\documentclass[a4paper,11pt]{article}
\pdfoutput=1 

\usepackage{jheppub} 
\usepackage[T1]{fontenc} 
\usepackage{slashed}
\usepackage[colorlinks=true,linktocpage=true]{hyperref}
\usepackage{cleveref}
\usepackage[numbers,sort&compress]{natbib}
\usepackage{mathtools}
\usepackage{xspace}
\usepackage{bm}
\usepackage{threeparttable}
\usepackage{xcolor}
\usepackage{graphicx} 
\usepackage{enumitem} 
\usepackage{listings}
\usepackage{hyperref}
\usepackage{fontawesome5}

\title{\boldmath The Bubble Wall Velocity in Local Thermal Equilibrium and Energy Budget with Full Effective Potential}

\author[a]{Zongguo Si,}
\author[a]{Hongxin Wang\footnote{Corresponding author},}
\author[b]{Lei Wang,}
\author[c,d,e]{Yang Xiao\footnote{Corresponding author},}
\author[c]{Yang Zhang}

\affiliation[a]{School of Physics, Shandong University, Jinan, Shandong 250100, China}
\affiliation[b]{Department of Physics, Yantai University, Yantai, Shandong 264005, China}
\affiliation[c]{School of Physics, Henan Normal University, Xinxiang 453007, China}
\affiliation[d]{Institute of Theoretical Physics, Chinese Academy of Sciences, Beijing 100190, China}
\affiliation[e]{School of Physical Sciences, University of Chinese Academy of Sciences,  Beijing 100049, P. R. China}

\emailAdd{zgsi@sdu.edu.cn}
\emailAdd{wanghx@mail.sdu.edu.cn}
\emailAdd{leiwang@ytu.edu.cn}
\emailAdd{xiaoyang@itp.ac.cn}
\emailAdd{zhangyang2025@htu.edu.cn}

\abstract{ 
We develop a framework based on the full one-loop finite-temperature effective potential model, within which the bubble wall velocity is calculated using the local thermal equilibrium (LTE) approximation, and the kinetic energy fraction $K$ is computed directly. In cosmological phase transitions, these quantities play a critical role in determining the resulting gravitational wave signals. Using the xSM as a benchmark model, we compute the peak gravitational wave spectra under different methods for determining the wall velocity and the kinetic energy fraction $K$, and compare these results to those obtained using the commonly employed bag model. Within the scanned parameter space, we find: (1) Deflagration is the most prevalent mode of fluid motion.(2) Gravitational wave spectra based on the full effective potential with LTE-derived wall velocity and integrated  $K$  can differ significantly from those using the bag model with fitted $K$. In the deflagration regime, discrepancies reach up to 48\% in peak frequency and 90\% in amplitude.
 (3) The bag model provides a good approximation to the full equation of state in many cases. Notably, in deflagration scenarios with input wall velocity, the gravitational wave spectra obtained from the bag model more closely resemble the LTE-based results than those derived using the full potential with this input wall velocity.
\href{https://github.com/bwlte2025/bubblewall_LTE}{\faGithub}
}

\begin{document} 
\maketitle
\flushbottom

\section{Introduction}
A first-order phase transition in the early universe is essential for realizing electroweak baryogenesis~\cite{Cohen:1993nk, Morrissey:2012db,Trodden:1998ym,Anderson:1991zb}, as it provides the necessary departure from thermal equilibrium required by the Sakharov conditions~\cite{sakharov1998violation}. However, the Standard Model (SM), constrained by the observed Higgs boson mass, cannot support a strongly first-order phase transition (FOPT). Thus, to achieve a viable FOPT, the SM must be extended—often through the introduction of new scalar fields or other degrees of freedom—leading naturally to scenarios beyond the Standard Model (BSM). These extensions not only enable baryogenesis but also provide the possibility of novel mechanisms for the production of dark matter~\cite{Baker:2018vos, Baker:2019ndr,Ayazi:2025sfs, Jiang:2023nkj,Xiao:2022oaq}. Because of its close connection to new particles, FOPT can impose additional constraints on Higgs decay channels, making it possible to indirectly probe the existence of FOPT at colliders~\cite{Profumo:2014opa, Contino:2016spe,Kotwal:2016tex,No:2013wsa,Profumo:2007wc,Bell:2020hnr,Zhang:2023jvh,Chen:2017qcz}. Meanwhile, FOPT can act as a source of stochastic gravitational waves, with a typical peak frequency in the millihertz range, which may be detected by future space-based detectors such as Taiji~\cite{Gong:2021any}, TianQin~\cite{TianQin:2015yph}, and LISA~\cite{LISA:2017pwj}. FOPT not only connects BSM physics with gravitational waves, but itself can also be simulated in condensed matter systems, offering a novel perspective on cosmological phase transitions~\cite{Braden:2017add, Billam:2020xna, Billam:2021psh}.

In a FOPT, the effective potential typically features two or more minima separated by potential barriers, corresponding to distinct possible vacuum expectation values (VEVs). As the transition begins, the local VEV of the scalar field undergoes a sudden change in a local region, leading to the nucleation of spherical structures known as critical bubbles~\cite{PhysRevD.15.2929}. Because of the pressure difference between the inside and outside of the bubble, it will expand outward. In the thermal plasma environment of the early universe, this expansion experiences friction from interactions with the surrounding particles. Thus, if the friction is sufficiently strong, the bubble wall will reach a terminal velocity; otherwise, the bubble enters a so-called "runaway" regime.Due to the plasma near the bubble, different steady-state bubble velocities lead to distinct fluid profiles. By matching the equation of state of the fluid to the bag model~\cite{Ai:2021kak} or the $\mu \nu$ model~\cite{Ai:2023see}, one can relate the resulting fluid profile to the phase transition strength $\alpha$ and the bubble wall velocity $\xi_w$. Clearly, different fluid profiles contribute differently to the gravitational wave spectrum, making the bubble wall velocity a crucial parameter in gravitational wave predictions. Moreover, the bubble wall velocity also affects the process of electroweak baryogenesis. While successful baryogenesis typically requires a relatively slow-moving wall to allow for sufficient CP-violating diffusion ahead of the bubble, gravitational wave generation favors a fast-moving wall to maximize the energy available for sourcing the signal. Reconciling these two competing requirements within a unified framework remains a significant theoretical challenge~\cite{Cline:2021iff, Beniwal:2018hyi, Zhou:2020irf,Ellis:2022lft}.

Accurately determining the terminal wall velocity is a highly nontrivial task, as it involves solving the coupled scalar-fluid Boltzmann equations under non-equilibrium conditions. Recently, the open-source software \texttt{WallGo} has been developed to solve the wall velocity, including those out-of-equilibrium contributions~\cite{Ekstedt:2024fyq}. In practice, it is common to assume a small deviation from local equilibrium and adopt an ansatz for the distribution functions to simplify the computation~\cite{Moore:1995ua,Moore:1995si,Jiang:2022btc}. In most phenomenological studies, the bubble wall velocity is treated as an input parameter to bypass these complexities. However, this simplification introduces significant uncertainty into subsequent results. For instance, in evaluating the energy budget of a FOPT, the fluid velocity profile near the bubble wall—classified as detonation, deflagration, or hybrid—depends sensitively on the wall velocity. These profiles critically determine the partitioning of vacuum energy; hence, arbitrary choices of bubble wall velocity can significantly affect the predicted energy budget and gravitational wave signals.

To minimize computational uncertainties, various alternative approaches have been developed~\cite{Ramsey-Musolf:2025jyk, Branchina:2025jou,Ai:2024btx,Yuwen:2024hme,Krajewski:2024gma,DeCurtis:2024hvh,Krajewski:2023clt,DeCurtis:2023hil,Laurent:2022jrs,Jiang:2022btc, DeCurtis:2022hlx,Wang:2020zlf,Megevand:2009gh,Ai:2025bjw,Ai:2023see,Ai:2021kak}, most of which aim to determine the steady-state bubble wall velocity by utilizing conservation laws. In Ref~\cite{Ai:2021kak}, a widely adopted method applies the entropy conservation condition under the local thermal equilibrium (LTE) approximation have been proposed, offering a fast and physically motivated estimate of the wall velocity. This method has been validated across several models, including the bag model, $\mu\nu$ model, singlet extends of SM (xSM)~\cite{Ai:2021kak,Ai:2024btx}, and the minimal left-right symmetric model (MLRSM)~\cite{Wang:2024wcs}. Furthermore, in the two limiting cases of collision rates—approaching zero and infinity—the lower and upper bounds of the wall velocity can be obtained using the ballistic and LTE approximations, respectively. Ref.~\cite{Krajewski:2024zxg} introduces a more general framework by assuming that the only source of entropy production during the FOPT is a dissipative, friction-like interaction between the thermal plasma and the expanding wall. This allows for a generalized matching condition beyond the conventional LTE approach, providing a complementary method for estimating the wall velocity.

In the aforementioned studies, the models considered are either simplified toy models or specific BSM scenarios, and thus lack a general framework that allows for the direct computation of the bubble wall velocity and fluid profiles from a complete effective potential. Developing such a model-independent approach is crucial for accurately predicting the gravitational wave signals from first-order phase transitions. In addition, with this framework in place, it becomes possible to investigate the impact of various approximate models—such as the bag model—on the energy budget, in comparison to the full effective potential. This, in turn, allows us to confirm the validity of these approximations. Therefore, in this work, we adopt the LTE approximation and aim to develop a general framework that enables users to estimate the bubble wall velocity and fluid profile based on the full effective potential, for any given set of BSM model parameters and fluid motion mode. Our approach offers an approximate estimate of the bubble wall velocity without requiring the explicit solution of the Boltzmann equations. Comparing with previous approaches where bubble wall velocity was chosen primarily to maximize the gravitational wave signal, our framework systematically constrains the wall velocity, thereby reducing the uncertainty in the gravitational wave results. 

This paper is organized as follows: In Section \ref{sec2}, we review the hydrodynamics of the FOPT and the LTE approximation. In Section \ref{sec3}, we illustrate our calculation method and validate our framework by comparing the results with those of the bag model. In Section \ref{sec4}, we perform a parameter space scan for the xSM with full effective potential and analyze the uncertainty in the resulting gravitational wave spectra. Finally, in Section \ref{sec5}, we draw the conclusions.

\section{Hydrodynamics and Local Thermal Equilibrium Approximation}\label{sec2}
\subsection{Equation of fluid motion and boundary conditions}
To derive the hydrodynamic description of the scalar field–plasma system during a FOPT, one begins with the covariant conservation of the total energy-momentum tensor in flat spacetime,
\begin{align}\label{eq:conservation of stress-energy}
  \partial_{\mu}T^{\mu\nu}=\partial_{\mu}(T^{\mu\nu}_\phi+T^{\mu\nu}_f)= 0.
\end{align}
Assuming the plasma is locally in thermal equilibrium and behaves as a perfect fluid, the energy-momentum tensors of the scalar field and the fluid are given by \cite{Espinosa:2010hh}
\begin{align}
    T^{\mu\nu}_\phi&=(\partial^\mu\phi)\partial^\nu\phi-g^{\mu\nu}\left(\frac{1}{2}(\partial\phi)^2-V_0(\phi)\right), \notag\\
    T^{\mu\nu}_f&=(\rho_{f}+p_{f})u^\mu u^\nu-p_{f} g^{\mu\nu},
\end{align}
where $V_0(\phi)$ is the zero-temperature effective potential, and $\rho_f$ and $p_f$ represent the energy density and pressure of the fluid, respectively. 
In the limit where the scalar field varies slowly and the system remains near LTE, the background field can be treated adiabatically. In this case, the fluid pressure can be written as $ p_{f}(\phi,T)= - V_T(\phi,T)$, where $V_T(\phi,T)$ denotes the thermal correction to the potential.
The total energy-momentum tensor takes the ideal fluid form 
\begin{align}
T^{\mu\nu}=(\rho+p)u^\mu u^\nu-p g^{\mu\nu},
\end{align}
where the pressure $p$ corresponds to the negative value of free energy density. 

Projecting the conservation equation of the energy-momentum tensor $\partial_{\mu} T^{\mu \nu} = 0$ onto the directions parallel and perpendicular to the fluid flow and assume that the solution is self-similar, which means the desired solution only depends on $\xi = r/t$, where $r$ is the distance from the bubble center and $t$ is the time elapsed since bubble nucleation. The hydrodynamic equations can then be written as
\begin{subequations}
\begin{align} 
\label{eq: diff equation_1}
   \frac{\left(\xi - v\right)}{w} \frac{\partial \rho}{\partial \xi} &= \frac{2v}{\xi} + \gamma^2 (1 - \xi v) \frac{\partial v}{\partial \xi},  \\
\label{eq: diff equation_2}
  \frac{\left(1- \xi v\right)}{w} \frac{\partial p}{\partial \xi} &= \gamma^2 (\xi -v) \frac{\partial v}{\partial \xi},
\end{align}
\end{subequations}
where $v(\xi)$ is the fluid velocity in the bubble center frame and $\gamma$ is the Lorenz factor and $\omega$ represents the total enthalpy, given by $\omega=p+\rho$. We can combine the two equations above into
\begin{equation} \label{eq: v_equation}
    \frac{2 v }{\xi} = \gamma^2 \left(1 - \xi v\right)  \left(\frac{\mu^2}{c_s^2}-1\right) \frac{\partial v}{\partial \xi},
\end{equation}
where $c_s$ is the speed of sound and
\begin{equation} \label{Lorentz velocity transformation}
    \mu(\xi, v) = \frac{\xi -v}{1 - \xi v}.
\end{equation}
When $\xi$ is set to the bubble wall velocity $\xi_w$, $\mu(\xi_w, v)$ will be the Lorentz velocity transformation between the bubble center frame and the bubble wall frame.
In addition, inserting Eq.~\eqref{eq: v_equation} into Eq.~\eqref{eq: diff equation_1} and Eq.~\eqref{eq: diff equation_2}, using the definition of the enthalpy density, we can obtain the below equation
\begin{equation} \label{eq: temperature_equation}
    \frac{\partial {\rm ln} T}{\partial \xi} = \gamma^2 \mu \frac{\partial v}{\partial \xi}.
\end{equation}
This equation means that the temperature across the wall is not a constant, but a profile closely related with the fluid velocity profile. Although this profile is often assumed to be constant, it can in fact influence the efficiency factor of sound wave-generated gravitational waves~\cite{Wang:2022lyd}.

To solve the above equation, a set of boundary conditions is required. Owing to the discontinuity in the VEV across the two sides of the bubble wall, the wall naturally constitutes a boundary interface separating distinct phases. To extract the matching conditions across the bubble wall, we consider a steady-state profile where the wall moves with constant velocity $\xi_w$. In the rest frame of the wall, all quantities depend only on the spatial coordinate perpendicular to the wall, and time derivatives vanish. Taking the direction of wall propagation as the positive $z$-axis, the fluid four-velocity takes the form 
\begin{align}
    u^\mu(z) = \gamma(z) \left(1,\, 0,\, 0,\,-v(z)\right).
\end{align}
In this setup, the non-trivial components of the conservation law \eqref{eq:conservation of stress-energy} reduce to $\partial_{z}T^{z\nu}= 0$, implying that both energy and momentum fluxes are conserved across the wall. This yield two matching condition
\begin{align}    
    \label{eq:matching condition1}
    &\omega\gamma^2 v={\rm const},\\
    \label{eq:matching condition2-1}
     &\omega\gamma^2 v^2+\frac{1}{2}(\partial_z\phi(z))^2+p={\rm const}.
\end{align}
These conditions provide the necessary boundary information to solve the scalar field and fluid profiles consistently in the wall frame~\cite{Espinosa:2010hh}.
The $zz$-component of the energy-momentum conservation equation yields the relation Eq.~\eqref{eq:matching condition2-1}.
Far away from the bubble wall, the scalar field reaches the minima of the effective potential and its spatial derivative vanishes, i.e., $\partial_z\phi(z)=0$. 
In this asymptotic region, the above condition Eq.~\eqref{eq:matching condition2-1} simplifies to
\begin{align} \label{eq:matching condition2}
\omega\gamma^2 v^2 + p = \text{const},
\end{align} 
and can be interpreted as the conservation of the longitudinal momentum flux across the wall.
Taking the difference of this expression between the symmetric and broken phases leads directly to the matching condition
\begin{align} 
\label{eq:pressure and back}
-\Delta p = \Delta{\omega \gamma^2 v^2}.
\end{align} 
From the perspective of bubble dynamics, Eq.~\eqref{eq:pressure and back} expresses a balance between the net driving force and the backreaction from the surrounding plasma. 
The left-hand side, $-\Delta p$ = $\Delta V_{\rm  eff}(\phi,T)$, represents the pressure difference across the bubble wall that drives its expansion. 
The right-hand side, $\Delta\{\omega\gamma^2v^2\}$, encodes the response of the external fluid—reflecting the resistance due to the inertia and motion of the plasma. 
This equation thus captures the dynamic equilibrium between the internal pressure pushing the wall outward and the opposing force exerted by the moving fluid.

Due to the balance between internal and external forces during the expansion of the bubble wall, the system evolves in a steady-state profile as the bubble wall propagates. That is, the motion of the bubble wall remains stable, without acceleration or deceleration. In such a regime, it is appropriate to assume local thermal equilibrium, under which the entropy current is conserved~\cite{Ai:2021kak}
\begin{align} 
\label{eq:entropy-divergence} 
\partial_\mu S^{\mu} \equiv \partial_\mu (s u^\mu) = 0. 
\end{align} 
Equation~\eqref{eq:entropy-divergence} further ensures the conservation of the total entropy for a fluid that remains static at spatial infinity.
To analyze the system within the rest frame of the bubble wall, we consider a steady-state profile where all fluid quantities, including entropy, temperature, and velocity, depend only on the spatial coordinate $z$. In this case, the conservation equation becomes 
\begin{align} 
\partial_z (s(z) \gamma(z)v(z)) = 0,
\end{align} 
implying this quantity is a constant across the wall. Combining with the thermodynamic relation $\omega=sT$, it allows us to write 
\begin{align} 
\omega \gamma^2 v = s T \gamma^2 v = \text{const} \Longrightarrow \gamma T = \text{const},
\end{align} 
where the entropy conservation condition have been used. The result is the third matching condition to solve the equations above, and also the extra condition beyond the bag model stated below. 

\subsection{Equation of state: Bag model}
From a thermodynamic perspective, the effective potential can be regarded as the free energy density 
and all thermodynamic state parameters can be directly computed via the below relations:
\begin{subequations}
\begin{align} 
    \label{eq: state parameter relation p}
    p &= -V_{\text{eff}}(\phi,T), \\
    \label{eq: state parameter relation e}
    \rho &= V_{\text{eff}}(\phi,T) - T \frac{\partial V_{\text{eff}}(\phi,T)}{\partial T}, \\
    \label{eq: state parameter relation w}
    w &= p + \rho = - T \frac{\partial V_{\text{eff}}(\phi,T)}{\partial T} ,\\
    s &= \frac{\omega}{T}= - \frac{\partial V_{\text{eff}}(\phi,T)}{\partial T}.
\end{align}
\end{subequations}
In general, the effective potential can be divided into the following components
\begin{align}\label{eq: full poential}
    V_{\rm eff}(\phi, T) = V_0(\phi) + V_{\rm CW}(\phi) + V_T(\phi, T),
\end{align}
where $V_0(\phi)$ is the tree level potential, $V_{\rm CW}(\phi)$ is the Coleman-Weinberg potential and $V_T(\phi, T)$ is the temperature dependent part and can be calculated by the finite temperature field theory. At one loop correction, $V_T$ can be expressed as 
\begin{equation}\label{eq: the temperature-dependent correction}
    V_T = \sum_{i=B,F} \pm g_i T \int \frac{d^3 k }{(2 \pi)^3} {\rm log}\left(1 \mp e^{-\sqrt{k^2 + m_i^2}/T} \right) = \sum_{i=B,F}\frac{T^4}{2 \pi} J_{B/F}\left(\frac{m_i}{T}\right),
\end{equation}
where $g_i$ is the relativistic degrees of freedom of each particle and the index "B"("F") is for bosons and fermis, respectively. In principle, the full effective potential should be used to solve the hydrodynamic equations. However, at high temperatures, a expansion can be employed to obtain an analytical expression for the temperature dependent part of the effective potential
\begin{align}
    \frac{T^4}{2 \pi}J_{B} &\approx -\frac{\pi^2}{90}T^4, \notag \\
    \frac{T^4}{2 \pi}J_{F} &\approx -\frac{7}{8}\frac{\pi^2}{90}T^4.
\end{align}
We can also neglect $V_{\rm CW}$, because it is corrections of $\mathcal{O}(g^4)$. Then, the effective potential can be simplified as a simple form
\begin{align} \label{eq: bag_V_approx}
    &V_{\rm eff}(\phi,T) = V_0(\phi) - \frac{1}{3}a T^4, \notag \\
    &a = \frac{\pi^2}{30} \sum_{i}\left(g_i^B + \frac{7}{8}g_i^F\right).
\end{align}
This is so-called bag model approximation. Physically, this approximation is justified because the masses of particles in the symmetric phase remain zero, making the expansion valid in that region. In the broken phase, the contributions from heavy particles are exponentially suppressed, so only lighter particles significantly affect the dynamics. The essence of the bag model lies in neglecting the contributions from particles whose masses are comparable to or exceed the temperature, thus simplifying the overall structure of the effective potential. A more refined simplification involves expanding the potential to higher orders, thereby introducing both quadratic and linear terms in temperature. More details can be found in Ref.~\cite{Wang:2022lyd}.

In bag model, the energy density $\rho$ and pressure $p$ of both the symmetric and broken phases are usually described by the equation of state below 
\begin{align} \label{eq: bag equation of state}
\rho_s &= a_{s}T^4 + \epsilon_{s}, ~~~~~~\rho_b = a_{b}T^4 + \epsilon_{b},\notag \\
p_s &= \frac{1}{3}a_{s} T^4 - \epsilon_{s}, ~~~~p_b = \frac{1}{3}a_{b} T^4 - \epsilon_{b},
\end{align}
where we have renamed $V_0$ as $\epsilon$ and the subscript $s$ represents the physical quantities in symmetric phase, while subscript $b$ denotes those in broken phase. $a_{s, b}$ is a dimensionless quantity, proportional to the degrees of freedom of particles in the symmetric and broken phases. Note that $a_{s} \ne a_{b}$ in the general case, as certain degrees of freedom are neglected due to the corresponding particles acquiring mass in the broken phase. The speed of sound is defined by below
\begin{align} \label{eq: speed of sound}
c^2_s=\frac{\partial p}{\partial \rho}
\end{align}
and it value is always $\frac{1}{3}$ for bag model in both phases. It is worth noting that this holds only when the pressure and energy density follow the relations given in Eq.~\eqref{eq: bag equation of state}. In more realistic models with interactions or massive particles, the speed of sound may deviate from this value, leading to a modification of the efficiency factor of sound wave~\cite{Wang:2021dwl, Giese:2020rtr, Wang:2020nzm}. Besides, because temperature is actually location dependent and thus the speed of sound is also location dependent rather than a simple constant.

To solve Eq.~\eqref{eq: v_equation}, the bag model assumes a known bubble wall velocity, allowing the system to be determined using only two boundary conditions mentioned above
\begin{subequations}
\begin{align}
    \label{matching condition1}
    w_{+}\bar{v}_{+}\gamma_{+}^2 &= w_{-}\bar{v}_{-}\gamma_{-}^2,  \\
    \label{matching condition2}
    w_{+}\bar{v}_{+}^2\gamma_{+}^2 + p_{+}&= w_{-}\bar{v}_{-}^2\gamma_{-}^2 + p_{-},
\end{align}
\end{subequations}
where the subscript "$+$" represents the physical quantities just ahead of the bubble wall, while subscript "$-$" denotes those just behind the bubble wall and $\bar{v}$ is the fluid velocity in the bubble wall frame. By introducing the strength factor near the front of bubble wall
\begin{equation}
    \alpha_{+} = \frac{\epsilon_{s} - \epsilon_{b}}{a_{+}T^4},
\end{equation}
the matching condition ~\eqref{matching condition1} and ~\eqref{matching condition2}  can be expressed as
\begin{align}
\label{matching condition in bag model}
       \bar{v}_{+}\bar{v}_{-} &= \frac{1 - \left(1 - 3 \alpha_{+}\right)r}{3 - 3\left(1 + \alpha_{+}\right)r}, \notag \\
       \frac{\bar{v}_{+}}{\bar{v}_{-}} &= \frac{3 + \left(1 - 3 \alpha_{+}\right)r}{1 + 3\left(1 + \alpha_{+}\right)r}, 
\end{align}
where $r = w_{+} /w_{-}$. 
Eq.~\eqref{matching condition in bag model} highlights a central feature of the bag model: once $\alpha_+$ and the fluid velocity on one side of the bubble wall are specified, the velocity on the other side can be uniquely determined via $r$.

Given a specified bubble wall velocity $\xi_w$, the fluid profile can be classified into three distinct modes:

{\bf (1) Detonation} ($\bar{v}_{+} >  \bar{v}_{-}$ and $ \bar{v}_{-} > c_s$):
A detonation profile propagates supersonically, and thus no shock wave forms ahead of the bubble wall. Instead, a rarefaction wave follows the wall. Since the fluid ahead of the supersonic bubble wall cannot anticipate its approach, it remains undisturbed. Therefore, in the comoving frame, the fluid in front of the wall is at rest relative to the bubble center:
\begin{equation}
        v_+ = 0.
\end{equation}
This leads to the following relations: 
\begin{equation}
\label{eq:detbnd}
	T_+ = T_N,\quad
	\bar{v}_{+}=\mu(\xi_w,v_+) = \xi_w.
\end{equation}
where $T_N$ is the nucleation temperature and this quantity can also be other reference temperature, like percolation temperature. In this case, the bubble wall velocity must exceed the Jouguet velocity $v_J$, which has a simple expression related with $\alpha_{+}$
\begin{align}
    v_J = \frac{\sqrt{\alpha_{+}(2 + 3 \alpha_{+}) + 1}}{\sqrt{3}(1 + \alpha_{+})}.
\end{align}

{\bf (2) Deflagration} ($ \bar{v}_{-} >\bar{v}_{+}$ and $ \bar{v}_{-} < c_s$):
A deflagration profile propagates subsonically, leading to the formation of a compression wave—commonly referred to as a shock wave—in the fluid ahead of the subsonic bubble wall. In the bubble center frame, the fluid behind the bubble wall remains at rest relative to the bubble center:
\begin{equation}
        v_- = 0,\quad
        \bar{v}_{-}=\mu(\xi_w,v_-) = \xi_w.
\end{equation}
In this case, the temperature $T_-$ behind the wall differs from the nucleation temperature $T_N$, while the temperature in front of the shock, denoted as $T_{sh,+}$, equals $T_N$.
The velocity profile of the compression wave exhibits a discontinuous drop to zero at the shock front, and the fluid ahead of the shock front is also at rest in the bubble center frame:
\begin{equation}
        T_{sh, +} = T_N,\quad
        v_{sh,+} = 0,\quad
        \bar{v}_{sh,+}=\mu(\xi_{sh},v_{sh,+}) = \xi_{sh}.
\end{equation}

{\bf (3) Hybrid} ($ \bar{v}_{-} > \bar{v}_{+}$ and $ \bar{v}_{-} = c_s$):
A hybrid profile generates both compression (shock) waves and rarefaction waves on both sides of the bubble wall. This type of solution emerges when the bubble wall velocity falls within the range between the speed of sound and the Jouguet velocity.
\begin{equation}
        \bar{v}_{-}=c_s,\quad
        {v}_{-}=\mu(\xi_w,\bar{v}_{-}).
\end{equation}
At the shock front, the conditions are the same as in the deflagration case.

In the cases of deflagration and hybrid modes, the presence of a shock wave prevents a direct calculation of the corresponding $\alpha_{+}$. Instead, by introducing $\alpha_{N}$, defined as the value of $\alpha$ far ahead of the bubble wall, we can relate it to $\alpha_{+}$ via the relation
\begin{equation}
    \frac{\alpha_{+}}{\alpha_{N}} = \frac{w_{N}}{w_+(\xi_w, \alpha_{+})}.
\end{equation}
 Here, $w_N$ denotes the enthalpy density in the symmetric phase far in front of the bubble wall, while 
$w_+(\xi_w, \alpha_{+})$ represents the enthalpy just in front of the wall at velocity $\xi_w$, expressed as a function of $\alpha_+$.
\section{Framework for full effective potential}\label{sec3}
As mentioned earlier, the necessity of manually inputting the bubble wall velocity in the bag model introduces significant uncertainties in subsequent gravitational wave predictions. To mitigate these uncertainties as much as possible, we aim to develop a more self-consistent computational framework by combining the full effective potential with the assumption of LTE. One can specify a fluid motion mode, and the framework will automatically determine whether such a profile is physically allowed for the given set of BSM model parameters. If it exists, the corresponding fluid profile will be returned.

The first major challenge in establishing such a framework lies in the complexity of the full effective potential, which makes it difficult to define a well-behaved parameter $\alpha_{+}$ that simplifies the boundary conditions into a similar form as Eq.~\eqref{matching condition in bag model}. Thus, we must rely on the original, unsimplified form of the boundary conditions Eq.~\eqref{matching condition1} and Eq.~\eqref{matching condition2} to determine the physical quantities near the bubble wall. By using the definition of enthalpy, those conditions can be written in a more convenient form
\begin{align} \label{eq: match condition}
    \bar{v}_{+}\bar{v}_{-} &= \frac{p_{+} - p_{-}}{\rho_{+} - \rho_{-}},\notag \\
    \frac{\bar{v}_{+}}{\bar{v}_{-}} &= \frac{\rho_{-} + p_{+}}{\rho_{+} + p_{-}}.
\end{align}
Both $p_{\pm}$ and $\rho_{\pm}$ depend explicitly on the VEV and the temperature. Since the VEV typically exhibits weak temperature dependence, it is a reasonable approximation to neglect this variation and fix the VEVs in the two phases to their values at a reference temperature. Under this assumption, Eq.~\eqref{eq: match condition} involves four unknowns, which can be chosen as $T_{\pm}$ and $\bar{v}_{\pm}$. Since there are only two independent equations, two of the unknowns must be specified in advance to ensure the system has a unique solution. In practice, this typically involves specifying the temperature and bubble wall velocity on one side of the bubble wall as input.

After introducing the third matching condition under the LET approximation, an additional constraint is added to the system, reducing the number of free variables. This means that only one input variable needs to be provided, allowing the remaining three variables to be solved, which makes it possible to numerically determine the bubble wall velocity.
We list the unknowns to be solved and the specified quantities that need to be provided for the three fluid motion modes, both before and after the introduction of the third matching condition, in Table.\ref{modes Unknowns}. For clarity, the specified quantities are highlighted in bold in the table. In Hybrid mode, the unknown $\bar{v}_{-}$ is omitted, because $\bar{v}_{-}$ is equal to the speed of sound in the broken phase and is a function of $T_-$ in this case.

\begin{table} 
\centering
	\begin{tabular}{|| c || c || c || c ||}
		\hline
		& Detonation & Deflagration & Hybrid    \\
		\hline
		\hline 
		$\textbf{two matching condition}$ 
        &~ $\bm{T}_{+}$,$\bm{\bar{v}_{+}}$,$T_-$,$\bar{v}_{-}$~
        &~$T_+$,$\bar{v}_{+}$,$\bm{T}_{-}$,$\bm{\bar{v}_{-}}$~     
        &~~ $T_+$,$\bar{v}_{+}$,$\bm{T}_{-}$~~         
         \\  
		\hline
		$\textbf{three matching condition}$ 
        & $\bm{T}_{+}$,$\bar{v}_{+}$,$T_-$,$\bar{v}_{-}$
        &$T_+$,$\bar{v}_{+}$,$\bm{T}_{-}$,$\bar{v}_{-}$    
        & $T_+$,$\bar{v}_{+}$,$T_- $      
          \\  
		\hline
	\end{tabular}
\caption{Unknowns and specified variables for different fluid motion modes before and after the introduction of the third matching condition. }
\label{modes Unknowns}
\end{table}

More specifically, all the three matching conditions can be equivalently written as
\begin{align}
\label{first matching cindition}
        \bar{v}_{+}^2&=\frac{p_{+} - p_{-}}{\rho_{+} - \rho_{-}} \times \frac{\rho_{-} + p_{+}}{\rho_{+} + p_{-}},\\
\label{second matching cindition}
        \bar{v}_{-}^2&=\frac{p_{+} - p_{-}}{\rho_{+} - \rho_{-}} /\frac{\rho_{-} + p_{+}}{\rho_{+} + p_{-}},\\
\label{third matching cindition of deflagration}   
  \frac{T_+}{T_-}&=\frac{\sqrt{1-\bar{v}_{+}^2(T_+,T_-)}}{\sqrt{1-\bar{v}_{-}^2(T_+,T_-)}}.
\end{align}

For detonation, the temperature in front of the bubble wall remains at the reference temperature $T_n$, as there is no shock wave present in this region. So we could first insert Eq.~\eqref{second matching cindition} into Eq.~\eqref{third matching cindition of deflagration}, expressing $\bar{v}_{+}^2$ as a function of $T_{+}$ and $T_{-}$. Then, $T_{-}$ can be determined by numerical methods. Once two parameters are known, solving for the remaining $\bar{v}_{-}$ and $\bar{v}_{+}$ becomes straightforward. Moreover, in this mode, the bubble wall velocity $\xi_w$ is simply given by $\bar{v}_{+}$.

For weak deflagration, the existence of a shock front introduces additional complexity into the solution process. To obtain the four variables, we first rewrite Eq.~\eqref{third matching cindition of deflagration} by substituting $\bar{v}_{-}$ (Eq.~\eqref{first matching cindition}) and $\bar{v}_{+}$ (Eq.~\eqref{second matching cindition}). Note that, because of the shock front, $T_+$ only denotes the temperature just ahead of the bubble wall, rather than the nucleation temperature $T_N$. After giving the guess value for $T_{-}$, we can determine $T_{+}$ by rewriting Eq.~\eqref{third matching cindition of deflagration} and then calculating the corresponding guess value for $\bar{v}_{+}$ and $\bar{v}_{-}$. The guessed bubble wall velocity can be given by $\xi_w$ = $\bar{v}_{-}$. The shock front is also a boundary and we can express its matching conditions as 
\begin{align} 
    \bar{v}_{sh, +}\bar{v}_{sh, -} &= \frac{p_{sh, +} - p_{sh, -}}{e_{sh, +} - e_{sh, -}}, \\
    \frac{\bar{v}_{sh, +}}{\bar{v}_{sh, -}} &= \frac{e_{sh, -} + p_{sh, +}}{e_{sh, +} + p_{sh, -}}.
\end{align}
For a corrected value of $T_{-}$, the resulting $\bar{v}_{sh, +}$ and $\bar{v}_{sh, -}$ must satisfy the above conditions. Thus, the $T_{+}$ and $v_{+}$ solved by a guessed $T_{-}$ could serve as initial conditions to solve Eq.~\eqref{eq: v_equation} and Eq.~\eqref{eq: temperature_equation}. For the resulting velocity profile, if there exist values of $\bar{v}_{sh, +}$ and $\bar{v}_{sh, -}$ that satisfy the above conditions, the guessed $T_{-}$ can be accepted. Otherwise, the guessed value of $T_{-}$ should be adjusted, and the procedure repeated. This iterative shooting method simultaneously determines the bubble wall velocity $\xi_w$ and the position of the shock front.

In the hybrid mode, the bubble wall velocity does not explicitly appear in the matching conditions, resulting in a solution procedure that differs from the previously discussed cases. A distinctive feature of this case is that $\bar{v}_{-} = c_s(T_-)$, which effectively reduces the number of independent variables by one. As a result, the remaining three unknowns across the wall can be determined by solving the three matching conditions. In particular, $T_+$ can be expressed in terms of $\bar{v}_{+}$ and $T_-$ as,
\begin{align} 
    T_+^2(\bar{v}_{+},T_-)=\frac{1-\bar{v}_{+}^2}{1-c_s^2(T_-)} T_-^2.
\end{align}
The solution procedure closely parallels that of the deflagration case. For a given range of guessed values for $\bar{v}_{+}$, the corresponding $T_{-}$, $\bar{v}_{-}$, and $T_{+}$ are determined using the two matching conditions at the bubble wall combined with a shooting method. To determine the bubble wall velocity $\xi_w$, we solve Eqs.~\eqref{eq: v_equation} and \eqref{eq: temperature_equation} using the obtained boundary conditions and a guessed value of $\xi_w$. The resulting fluid profile is then validated by checking whether the matching conditions at the shock front are satisfied. This procedure is iteratively repeated until convergence is achieved.

Up to this point, our analysis has not involved the parameter $\alpha_+$. Technically, within the framework of the bag model, it merely serves as an auxiliary quantity relating $\bar{v}_{+}$ and $\bar{v}_{-}$. This observation implies that one can bypass the definition of the strength factor $\alpha_+$ and instead construct the fluid configuration directly from the full effective potential by solving the hydrodynamic equations in conjunction with the matching conditions. Moreover, this approach simultaneously yields the associated temperature profile, providing access to more detailed and informative descriptions of the underlying fluid dynamics.

The second challenge lies in determining whether the assumed fluid motion mode is consistent with the obtained bubble wall velocity. In the bag model, this is typically assessed by comparing the wall velocity with the sound speed and the Jouguet velocity. While the speed of sound can be directly computed from a full effective potential, the determination of the Jouguet velocity is less straightforward. In Ref.~\cite{Wang:2024wcs}, the Jouguet velocity is treated as the minimum value of $\bar{v}_{+}$ for detonation that satisfies the below hydrodynamic and matching conditions
\begin{equation}
    v_J=\bar{v}_{+}^{min}(T_+=T_N,T_-,\bar{v}_{+}).
\end{equation}
Evidently, when the bubble wall velocity $\xi_w$ is smaller than the Jouguet velocity $v_J$, the corresponding  $\bar{v}_{+}$ is also less than $v_J$, indicating the absence of physical solutions in the detonation regime. This behavior is consistent with the definition of the Jouguet velocity in the bag model. We adopted this definition and have numerically verified that, within the bag model framework, the values of $v_J$ obtained from both definitions are in close agreement.

\begin{figure}[htbp!]
    \centering
    \includegraphics[page=1, width=0.8\textwidth]{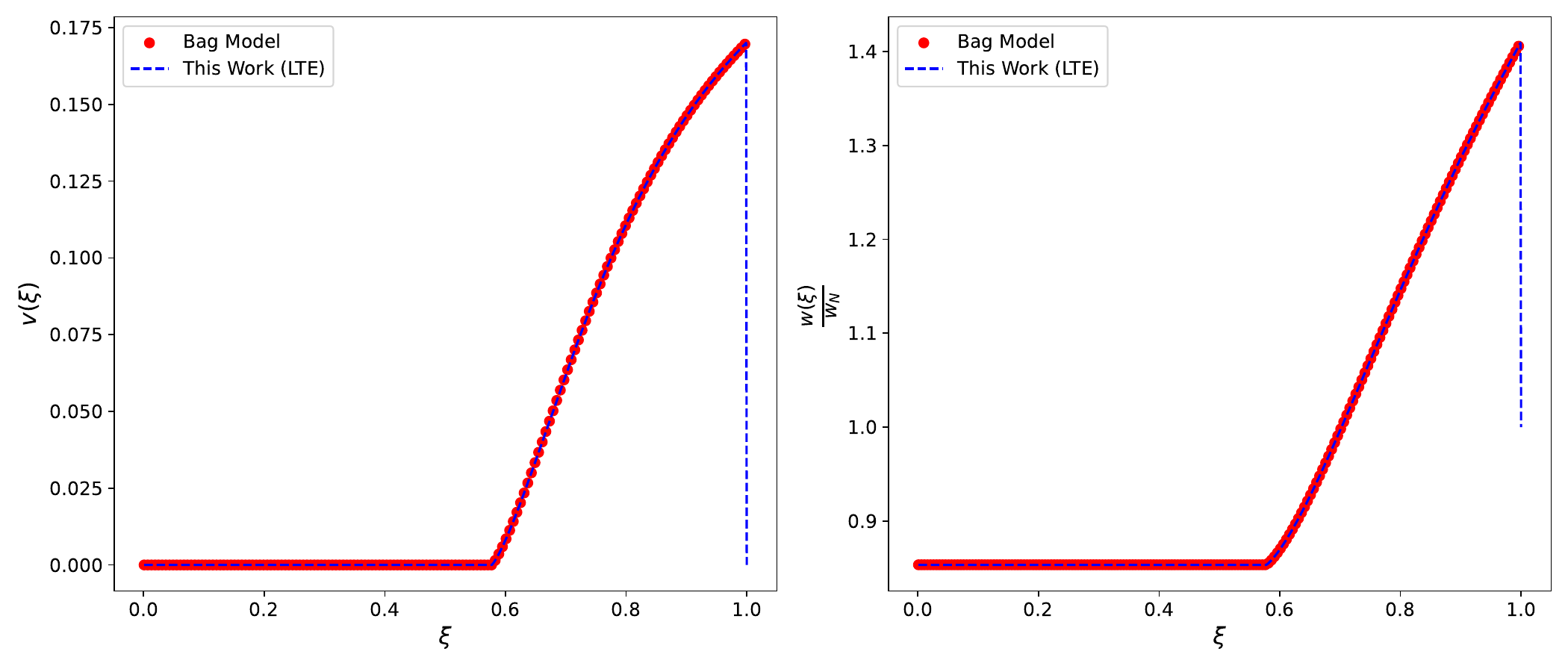}\\
    \includegraphics[page=1, width=0.8\textwidth]{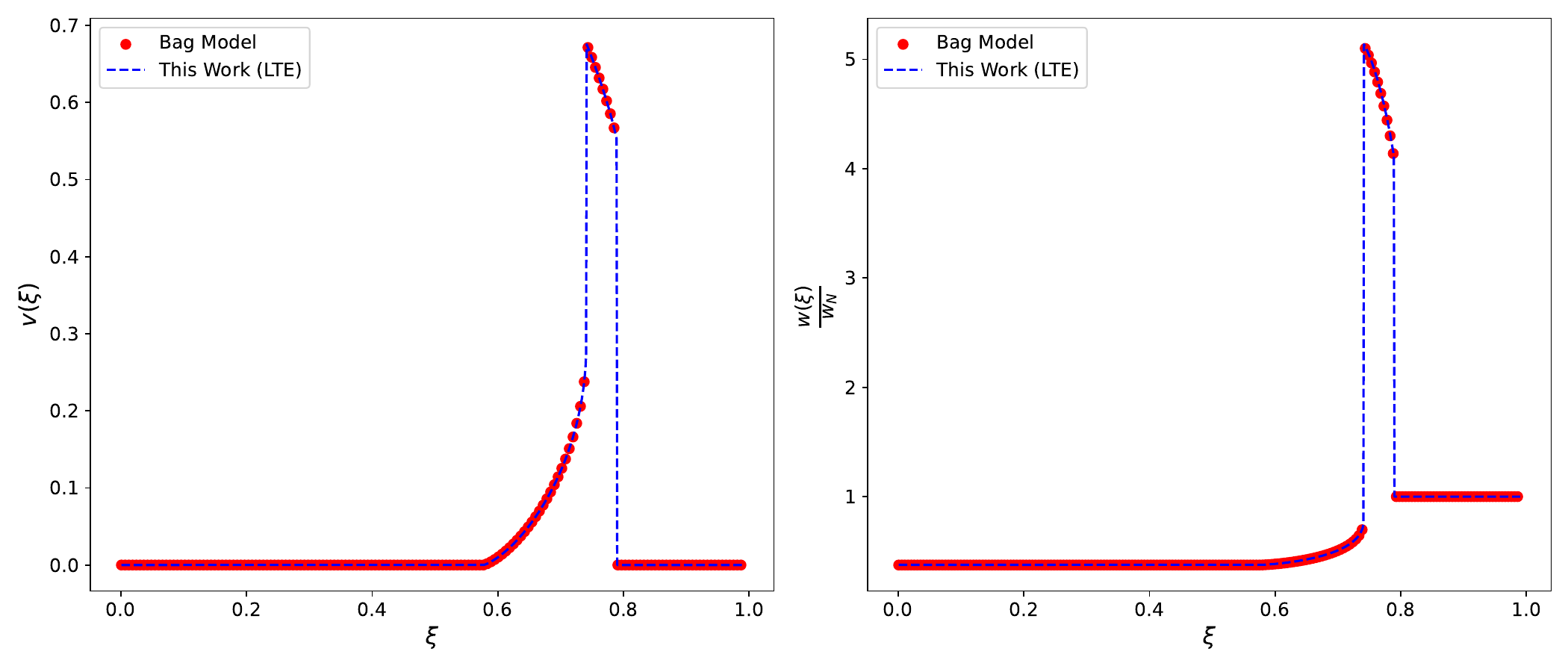}\\
    \includegraphics[page=1, width=0.8\textwidth]{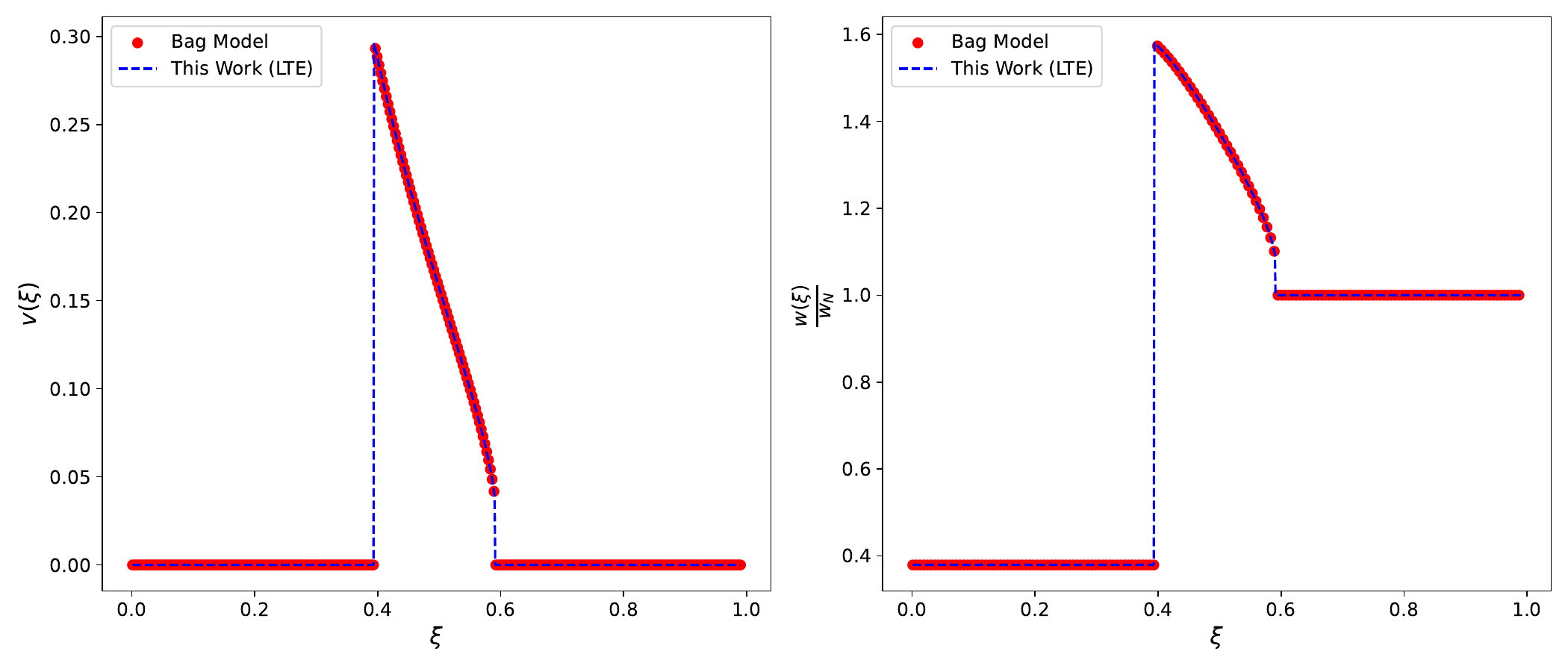}
    \caption{Comparison between our method and the previous approach, in which the auxiliary parameter $\alpha_{+}$ was introduced, within the artificially constructed bag model. The blue curve represents the results obtained using our method, while the red dots denote the outcomes from the previous approach.}
    \label{fig: 1}
\end{figure}

To validate our computational framework, we performed a benchmark test using an artificial bag model. Specifically, we chose the zero-temperature effective potential to be:
\begin{equation}
    V_0 = \frac{1}{2}M_s\phi^2 - \frac{\delta}{3}\phi^3 + \frac{\lambda}{4}\phi^4,
\end{equation}
and set $M_s=62.5$, $\delta = 110$ and $\lambda = 0.2$. The degree of freedom parameter $a_{+}$ and $a_{-}$ is set as 106.75 and 35.00, respectively. In the bag model framework, the fluid profile depends only on two input parameters: $\alpha_+$ and the bubble wall velocity $\xi_w$. Therefore, we first use our method to compute the bubble wall velocity for each given mode and then use $T_{+}$ to calculate the corresponding $\alpha_+$ for each mode. Using the resulting input pairs $(\alpha_+,\xi_w)$, we can obtain the corresponding fluid profiles within the bag model and compare them with previous obtained from our approach, which does not rely on $\alpha_+$. The results are presented in FIG.\ref{fig: 1}.

The left panel shows the fluid velocity profile, while the right panel displays the enthalpy profile. The blue dashed lines represent the results obtained directly from the effective potential under the assumption of local thermal equilibrium, without introducing the auxiliary parameter $\alpha_+$. The bubble wall velocities calculated for the three modes are 0.99, 0.74, and 0.39, respectively. The red dots indicate the results computed using the bag model.
As shown in the figure, the two results agree remarkably well across all three propagation modes, with excellent matching in both the location and magnitude of the velocity and enthalpy. This consistency demonstrates that our approach—despite not relying on the intermediate parameter $\alpha_+$ can reliably capture the essential features of the fluid profile. This confirms the validity and accuracy of our method.

In detonation mode, the calculated bubble wall velocity approaches the runaway limit. As discussed in Ref.~\cite{Ai:2021kak}, steady-state detonation solutions with wall velocities approaching the speed of light are nonphysical. This is because the wall velocity increases as dissipative friction decreases, causing the bubbles to become luminal before the dissipative force vanishes. Although such behavior is known to be sensitive to model parameters, the primary aim of this section is to validate the numerical methods by comparing the results with those obtained from the artificial model. Therefore, we did not fine-tune the parameters to achieve a more representative detonation solution with a wall velocity below 0.99. In subsequent analyses using the realistic model, detonation solutions with wall velocities approaching the speed of light were also observed. In line with the argument in Ref.~\cite{Ai:2021kak}, solutions with $\xi_w>0.99$ are regarded as nonphysical and discarded.

\section{Test BSM Model -- singlet extension of SM}\label{sec4}
In xSM, the full effective potential with loop corrections is given by~\cite{Athron:2022jyi}
\begin{equation} \label{eq: V_eff}
V_{\rm eff}(h, s; T) = V_{0}(h, s) + V_{\rm CW}(h, s)+  V_{1T}(h, s; T) + V_{\rm ring}(h, s; T), 
\end{equation}
where $V_{\rm CW}$, $V_{1T}$ and  $V_{\rm ring}$ represent the one-loop Coleman-Weinberg potential, the one-loop thermal correction, and the resummed daisy correction, respectively. We adopt $Z_2$ symmetry in tree level potential, so it takes the form 
\begin{equation}
V_0(h,s) = -\frac{\mu_h^2}{2} h^2 + \frac{\lambda_{h}}{4}h^4 - \frac{\mu_s^2}{2} s^2 + \frac{\lambda_s}{4} s^4 + \frac{\lambda_{hs}}{4} h^2 s^2 .
\label{Eq:tree}
\end{equation}
The OS-like scheme and the Landau gauge are chosen to avoid introducing dependence on the renormalization scale, 
and the corresponding one-loop zero-temperature correction is given by
\begin{equation}
\label{eq: V_CW}
\begin{aligned}
 V_{\rm{CW}} (h,s) 
 =&\frac{1}{64 \pi^2} \sum_i (-1)^{2 s_i} n_i \left[m_i^4(h,s) \left(\log\left(\frac{m_i^2(h,s)}{{m_i^2(h,0)}}\right) - \frac{3}{2} \right) + 2 {m_i^2(h,s)} {m_i^2(h, 0)} \right],
\end{aligned}
\end{equation}
where $i$ ranges over the entire field-dependent mass spectrum $m_i(h,s)$, including scalars ($h,s$), gauge bosons ($W,Z$) and fermion ($t$). To avoid divergence at zero temperature, we neglect the contribution of Goldstone bosons, which will not qualitatively affect our results. For a more refined treatment, please refer to Ref.~\cite{Elias-Miro:2014pca,Curtin:2016urg}. Finally, the one-loop finite-temperature correction $V_{1T}$ is introduced as in Eq.~\eqref{eq: the temperature-dependent correction}.

Due to contribution of Matsubara zero-mode, multi-loop corrections will dominate at high temperatures, which may break the perturbation condition. To ensure the reliability of the expansion, the dominant thermal contributions must be resummed. We adopt the Parwani method by replacing the tree-level masses $m_{i}^2$, with the thermal masses $m_{i}^2(T) = m_{i}^2 + d_{i}T^2$~\cite{Parwani:1991gq}, where $d_{i}T^2$ is the leading contribution in temperature to the one-loop thermal mass
\begin{equation} \label{eq: deby mass correction}
\begin{aligned}
d^{L}_{W^{\pm,3}}&=\frac{11}{6}g_{SU(2)_L}^2, ~~~~~~
d^{T}_{W^{\pm,3}}=0, ~~~~~~
d_{B}^{L}=\frac{11}{6}g_{U(1)_Y}^{2},\\
d_{HH} &= \frac{3g_{SU(2)_L}^2}{16} + \frac{1}{16}g_{U(1)_Y}^2+\frac{1}{2}\lambda_{H}+\frac{1}{4}y_{t}^2+\frac{1}{24}\lambda_{hs},\\
d_{SS} &= \frac{1}{4}\lambda_{s} + \frac{1}{6}\lambda_{hs}.
\end{aligned}
\end{equation}

Before adopting xSM as a template model for BSM studies, we first assess the accuracy of our numerical results by comparing the bubble wall velocity and the Jouguet velocity under the assumption of LTE with \texttt{WallGo}~\cite{Ekstedt:2024fyq}, using the benchmark parameters: $m_s = 120~\mathrm{GeV}$, $\lambda_{hs} = 0.9$ and $\lambda_s = 1.0$. For consistency, we reconstructed the finite-temperature effective potential using the publicly available code \texttt{CosmoTransitions}~\cite{Wainwright:2011kj}, consistent with the setup in Ref.\cite{Ekstedt:2024fyq}. The reconstructed xSM potential differs from the one discussed above; specifically, the potential $V_{\rm CW}$ adopts the $\overline{\text{MS}}$ renormalization scheme and neglects the Daisy resummation terms. In Ref.\cite{Ekstedt:2024fyq}, the nucleation temperature for this parameter is $T_N=100$ GeV, with the true and false vacuum configurations located at (195.032 GeV, 0.0 GeV) and (0.0 GeV, 104.868 GeV), respectively. In our reconstructed xSM potential,
the corresponding vacuum field values are (195.028 GeV, 0.0 GeV) and (0.0 GeV, 104.869 GeV). Although these values are not exactly identical to those reported in the reference, the discrepancies are minimal and can be attributed to numerical precision. So, we consider the phase transition histories described by \texttt{WallGo} and our reconstruction to be effectively equivalent. \texttt{WallGo} identifies this parameter point as corresponding to a hybrid propagation mode and yields a bubble wall velocity of 0.6203. Using the hybrid mode for consistency, our method yields a slightly lower value of 0.6127, resulting in a relative difference of approximately 1.22\%. The computed values of the Jouguet velocity are 0.6493 (ours) and 0.6442 (\texttt{WallGo}), corresponding to a relative deviation of about 0.79\%. The agreement between the two results validates the reliability of our computational approach.

\begin{figure}[tb]
	\centering
        \includegraphics[width=0.49\textwidth]{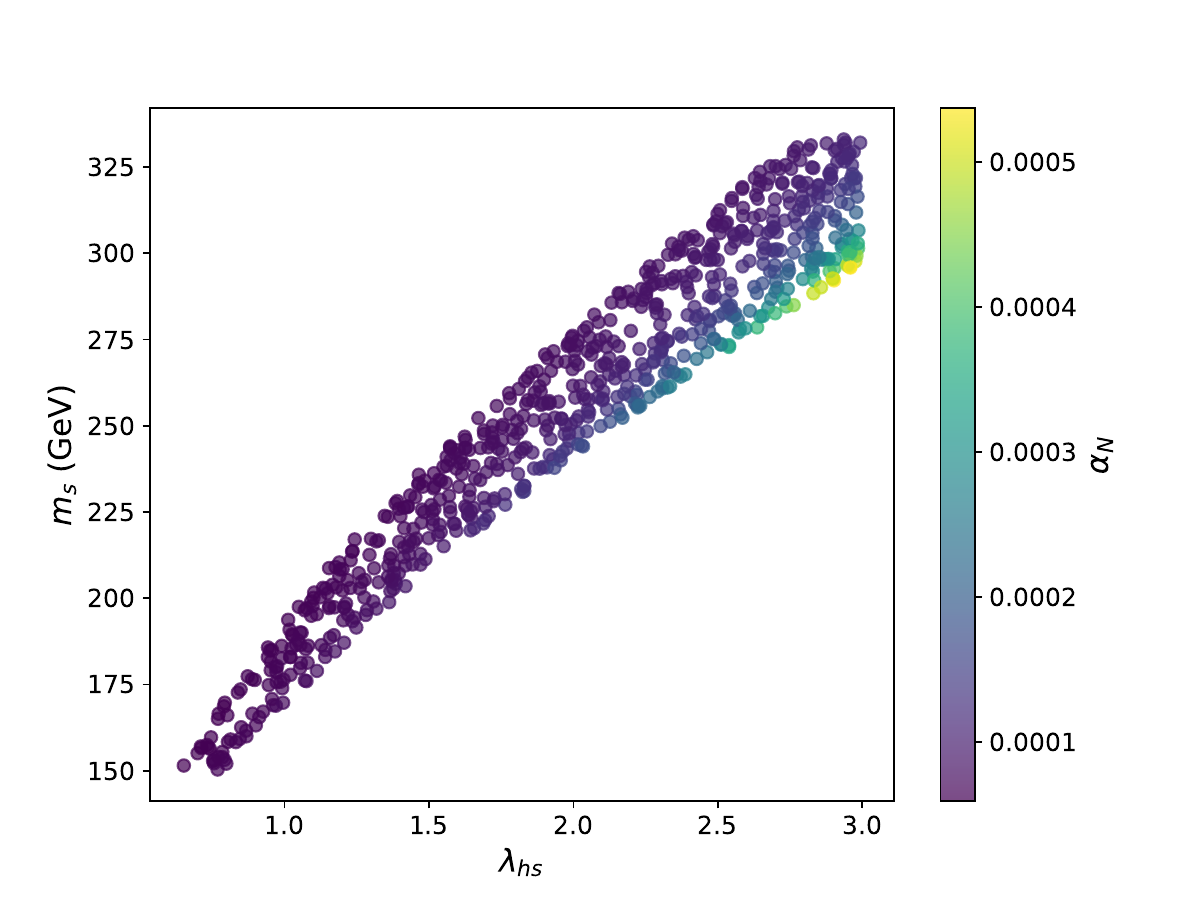}
        \includegraphics[width=0.49\textwidth]{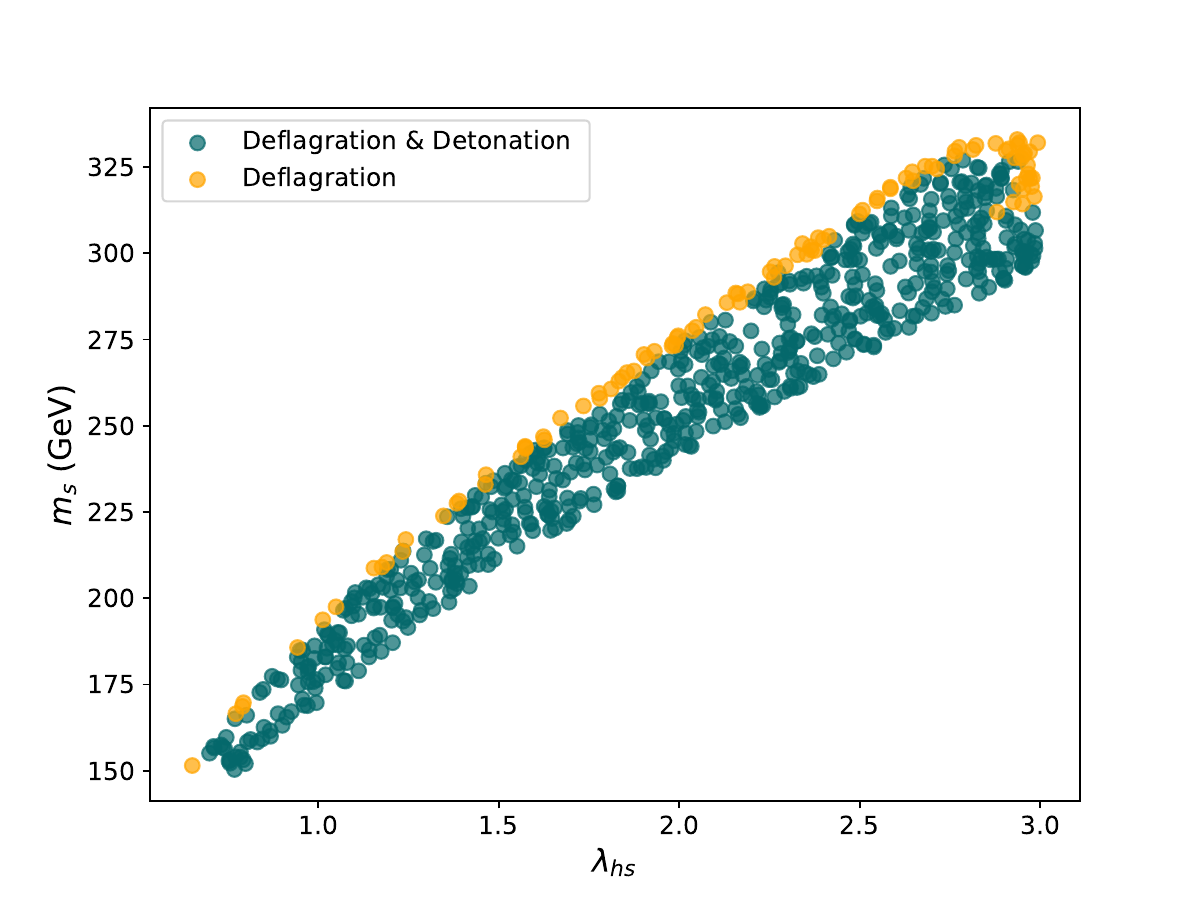}
	  \caption{Scan results for the singlet-extension of SM. (Left panel): Phase transition strength factor $\alpha_N$ as a function of the scanned parameters. (Right panel): Allowed hydrodynamic modes for each parameter point under the LTE approximation. Parameter points leading to runaway solutions (i.e., $\xi_w > 0.99$) have been excluded. The yellow and green points respectively indicate regions with only deflagration modes and those where both deflagration and detonation solutions coexist.}
	\label{Scan results for the singlet-extension of SM}
\end{figure}

We performed a parameter scan in the following ranges,
\begin{equation} \label{eq: scan_area}
    150 \mathrm{GeV} \le m_{s} \le 500 \mathrm{GeV}, ~~0 \le \lambda_{hs} \le 3, ~~\lambda_{s}=0.2
\end{equation}
with the results shown in Fig.\ref{Scan results for the singlet-extension of SM}. In the left panel, it can be observed that all parameter points for which fluid profiles are obtained correspond to relatively small values of $\alpha_N$. This is because, for larger $\alpha_N$, the bubble wall velocity predicted by the detonation mode tends to approach the speed of light and is thus discarded as unphysical. This observation indirectly supports the conclusion that the LTE approximation is valid only for small $\alpha_N$. In the right panel, the yellow points correspond to regions where only the deflagration modes are realized, while the green points indicate parameter regimes in which only detonation and deflagration solutions are found.  

As shown in the figure, deflagration emerges as the most readily realized steady-state solution. The formation of a shock wave leads to plasma accumulation ahead of the bubble wall, increasing the pressure in front of the wall and thereby suppressing its further acceleration. 
In scenarios where only deflagration solutions exist, the corresponding detonation solutions typically enter a runaway regime and therefore fail to satisfy the steady-state condition. These deflagration points are predominantly located near the boundaries of the parameter space. A comparison with the left panel reveals that these regions are characterized by relatively weak phase transitions $\alpha_N$. This behavior can be understood from the fact that the net pressure exerted on the bubble wall reaches its maximum at the Jouguet velocity $v_J$. When $\alpha_N$ is small, the wall lacks sufficient driving force to overcome this pressure barrier, thereby preventing the existence of a viable detonation solution.
However, prior to the establishment of a steady-state configuration, if the formation of the shock wave is sufficiently delayed, the wall may accelerate beyond $v_J$, making a detonation solution viable.

In the parameter space we consider, hybrid-mode solutions do not appear. There are two reasons. First, we define the Jouguet velocity $v_J$ as the minimum value of $\bar{v}_+$ in the detonation mode, where $\bar{v}_+$ corresponds to the intersection point of the two branches described by Eq.~(\ref{vplus two branch}),
\begin{align}
\label{vplus two branch}
       \bar{v}_{+} &= \frac{1 - \left(1 - 3 \alpha_{+}\right)r}{3 - 3\left(1 + \alpha_{+}\right)r}/\bar{v}_{-}, \notag \\
       \bar{v}_{+} &= \frac{3 + \left(1 - 3 \alpha_{+}\right)r}{1 + 3\left(1 + \alpha_{+}\right)r} \bar{v}_{-}, 
\end{align}
in certain regions of the parameter space, such an intersection does not exist, making it impossible to determine $v_J$. As the determination of the bubble wall velocity in the hybrid mode requires a well-defined $v_J$, hybrid solutions are not meaningful in scenarios where $v_J$ is ill-defined. These situations primarily occur in regions where only deflagration solutions exist. Second, in some scenarios, the fluid velocity profile computed under the hybrid assumption lacks a tail structure near the bubble wall—that is, it satisfies $\xi_w = c_s = \bar{v}_-$. This condition implies that the wall velocity equals both the speed of sound and the fluid velocity behind the wall, thereby preventing the formation of the supersonic transition region required for a hybrid profile. Such cases predominantly occur in regions where both deflagration and detonation branches coexist. It is worth emphasizing that although hybrid-mode solutions do not appear in the parameter space we scanned, this does not indicate any issue with the computation. Rather, it reflects the strong parameter dependence of the existence of hybrid solutions. In fact, this has been confirmed through the cross-check with \texttt{Wallgo} presented at the beginning of this chapter.

We now examine the uncertainties in the gravitational wave prediction within the xSM framework, with particular attention to those arising from the bubble wall velocity dependence in the sound wave contribution, as well as from the use of the bag model as an approximate equation of state (EOS). The contribution from sound waves is given in Ref.~\cite{Hindmarsh:2015qta} as
\begin{align}
\Omega_{\textrm{sw}}h^{2} &=
2.65\times10^{-6}\left( \frac{H(T_{*})}{\beta}\right)K^{2}
\left( \frac{100}{g_{\ast}(T_{*})}\right)^{1/3} {\xi}_w~\Upsilon(\tau_{\rm sw})\nonumber \\
&\times  \left(\frac{f}{f_{\rm sw}} \right)^{3} \left( \frac{7}{4+3(f/f_{\textrm{sw}})^{2}} \right) ^{7/2} \ ,
\label{eq:soundwaves}
\end{align}
where the peak frequency is given by
\begin{align}
f_{\textrm{sw}} \ = \ 
1.9\times10^{-5}\frac{1}{{\xi}_w}\left(\frac{\beta}{H(T_{*})} \right) \left( \frac{T_{*}}{100\textrm{GeV}} \right) \left( \frac{g_{\ast}(T_*)}{100}\right)^{1/6} \textrm{Hz}. \
\end{align}
Here, $\beta$ characterizes the inverse duration of the phase transition and is defined as,
\begin{equation}
\beta=TH(T)\frac{d (S_3/T)}{d T}|_{T=T_*}\; .
\end{equation}
with $H(T_{*})$ denoting the Hubble parameter at the reference temperature $T_{*}$, and $g_*(T_*)$ the number of relativistic degrees of freedom at that temperature.

The suppression factor $\Upsilon(\tau_{\rm sw})$, accounting for the finite lifetime of the sound wave source, is given by \cite{Guo:2020grp}.
\begin{equation}
\Upsilon(\tau_{\rm sw})=\left.1-\frac{1}{\sqrt{1+2\tau_{\rm sw}H(T)}} \right|_{T=T_*}\ ,
\end{equation}
where the lifetime $\tau_{\rm sw}$ is estimated as \cite{Ellis:2020awk,Wang:2020jrd},
\begin{equation}
\tau_{\rm sw}=\frac{{\xi}_w(8\pi)^{1/3}}{\beta \bar{U}_f}, ~~ \bar{U}^2_f=\frac{3}{4}K\ .
\end{equation}
The kinetic energy fraction $K$ represents the fraction of total energy transferred to the bulk motion of the plasma and is defined as
\begin{align}
K=\frac{\rho_{\rm kin}}{\rho_{\rm tot}},
\label{eq:the kinetic energy fraction}
\end{align}
where $\rho_{\rm kin}$ can be calculated from the enthalpy and velocity profiles obtained by solving the hydrodynamic equations,
\begin{align}
\rho_{\rm kin}=\frac{3}{\xi_w^3}\int\xi^2v^2\gamma^2wd\xi.
\end{align}
We choose the nucleation temperature $T_n$ as the reference temperature, and $\rho_{\rm tot}$ denotes the total energy density, $\rho_{\rm tot}(T) = \rho_f(T) - \rho_{\rm gs}$~\cite{Athron:2023xlk}, where $\rho_f$ is the energy density of false vacuum and $\rho_{\rm gs}$ is the energy density corresponds to the lowest phase at zero temperature. Alternatively, $K$ in bag model can also be rewritten using the transition strength $\alpha_N$ and the efficiency factor $\kappa_{\rm sw}$,  which quantifies the fraction of vacuum energy converted into kinetic energy,
\begin{align}
K=\frac{\kappa_{\rm sw}\alpha_N}{1+\alpha_N}.
\label{eq:the kinetic energy fraction fit}
\end{align}
The efficiency factor $\kappa_{\rm sw}(\xi_w,\alpha_N)$ depends on the bubble wall velocity and the phase transition strength, varying across different hydrodynamic regimes. Its numerical fitting results obtained from bag model can be described by the following expressions:
\begin{itemize}
    \item For the deflagration mode ($\xi_w \lesssim c_s$):
    \begin{align} \label{eq: fitting deflagration}
    \kappa_{\textrm{sw}} = \frac{c_s^{11/5} \kappa_A \kappa_B}{(c_s^{11/5} - \xi_w^{11/5}) \kappa_B + \xi_w c_s^{6/5} \kappa_A} ,
    \end{align} 
    \item For the hybrid mode ($c_s < \xi_w < v_J$)
    \begin{align} \label{eq: fitting hybrid}
    \kappa_{\textrm{sw}} = \kappa_B + (\xi_w - c_s) \delta k + \left( \frac{\xi_w - c_s}{v_J - c_s} \right)^3 \left[ \kappa_C - \kappa_B - (v_J - c_s)\delta k \right] ,
    \end{align}
    \item For the detonation mode ($v_J \lesssim \xi_w$)
    \begin{align} \label{eq: fitting detonation}
    \kappa_{\textrm{sw}} = \frac{(v_J - 1)^3 v_J^{5/2} \xi_w^{-5/2} \kappa_C \kappa_D }{[(v_J -1)^3 - (\xi_w - 1)^3] v_J^{5/2} \kappa_C + (\xi_w - 1)^3 \kappa_D} .    \end{align}
    
\end{itemize}
In the above, $\kappa_{A}(\xi_w,\alpha_N)$ and $\kappa_{B,C,D}(\alpha_N)$ are fitting functions that correspond respectively to the non-relativistic, sonic (sound-speed), Jouguet, and ultra-relativistic regimes.The term $\delta k(\alpha_N)$ represents the slope of $\kappa_{\rm sw}$ during the smooth transition between deflagration and hybrid modes. Detailed expressions for $\kappa_{A,B,C,D}$ and $\delta k$ can be found in Ref.~\cite{Espinosa:2010hh}.

\begin{figure}
    \centering
    \includegraphics[width=\linewidth]{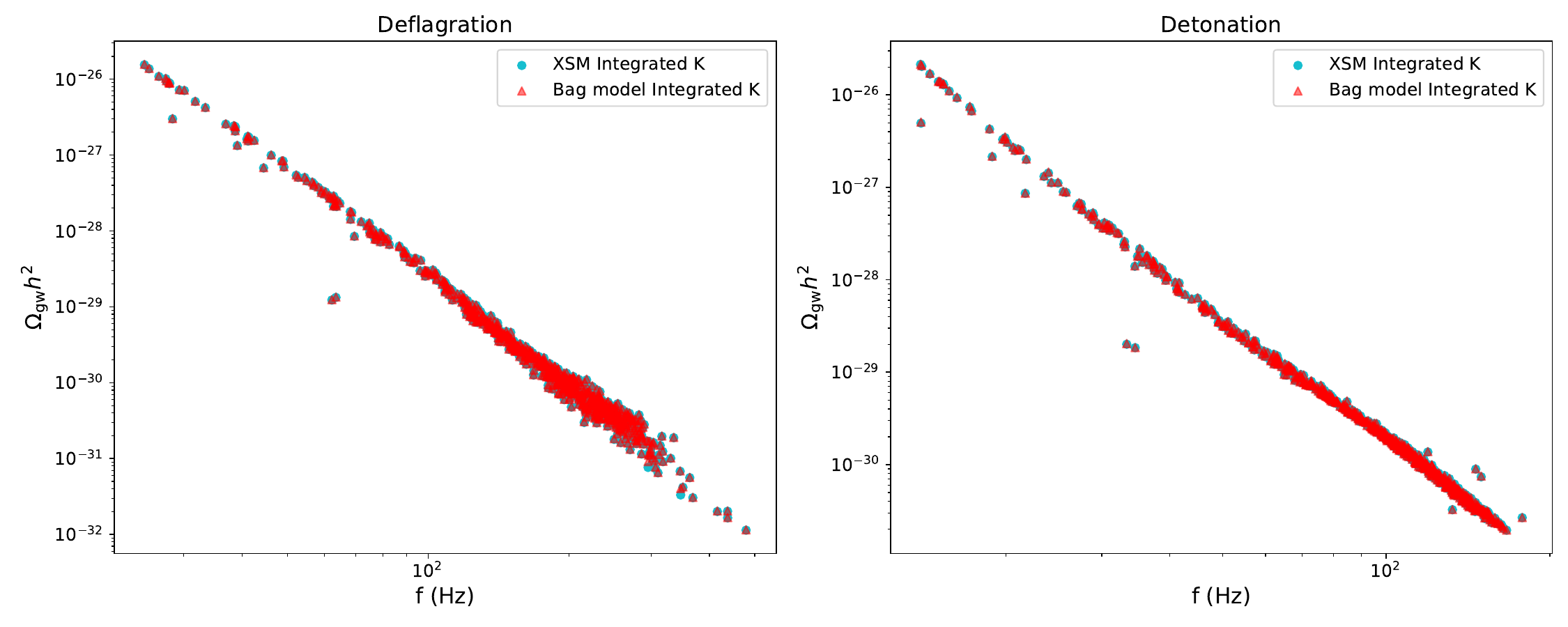}
     \includegraphics[width=\linewidth]{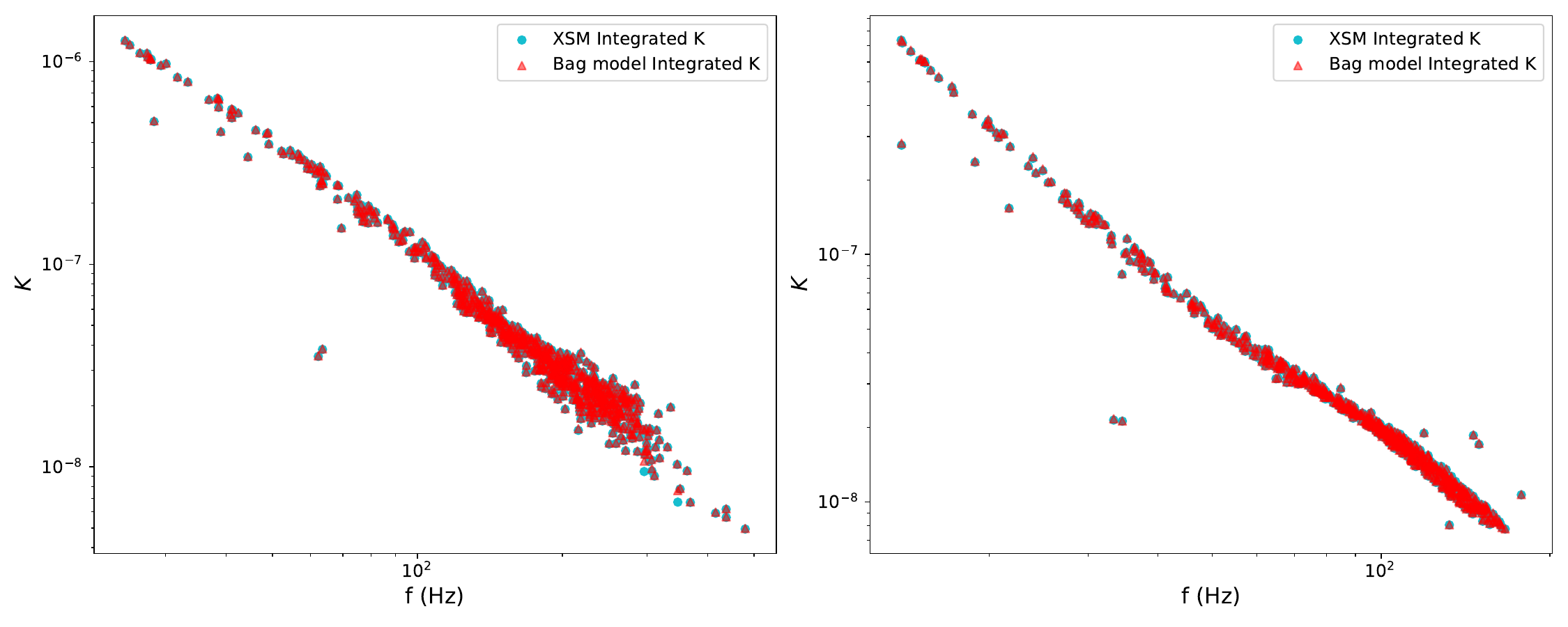}
    \caption{Comparison of gravitational wave peak spectra and energy budgets obtained with fixed bubble wall velocity (LET value) and a fixed method for calculating $K$ under different equations of state. The left panel corresponds to deflagration modes, and the right panel to detonation modes. The red triangles represent results calculated using the bag model, while the cyan circular data points correspond to those computed with the full effective potential of the xSM.}
    \label{fig:different_EOS_K_Omega}
\end{figure}

Given that the gravitational wave spectrum is sensitive to the EOS, the kinetic energy fraction $K$, and the bubble wall velocity $\xi_w$, we assess the uncertainties by varying each of these quantities individually, holding the other two constant in each case. 
First, we vary the EOS while $\xi_w$ is set to the LET result based on the full effective potential, and $K$ is computed using Eq.~(\ref{eq:the kinetic energy fraction}). The resulting spectra are shown in Fig.\ref{fig:different_EOS_K_Omega}. The cyan circular points represent the results obtained using the full EOS, while the red triangles correspond to the results computed with the bag model.

In the upper panel of Fig.\ref{fig:different_EOS_K_Omega}, we observe that the peak amplitudes of the gravitational wave spectra are nearly identical across the parameter space for both fluid motion profiles, with only minor deviations at the lower-right corner of the deflagration parameter region. 
According to the gravitational wave fitting formula Eq.(\ref{eq:soundwaves}), the peak amplitude is primarily determined by $K$ and $\xi_w$, where $K$ is mainly determined by the fluid profile. Ref.~\cite{Giese:2020znk} points out that substantial differences in the resulting fluid profile arise only when the speed of sound deviates significantly from the bag model value. We select several benchmark points and find that, under the full EOS, the squared speed of sound in the phases on both sides of the bubble wall is close to 1/3, consistent with the assumption made in the bag model. 
Therefore, the values of $K$ computed from the two different EOS should be nearly identical. We present the values of $K$ in the lower panel of Fig.\ref{fig:different_EOS_K_Omega}, which confirms this expectation. This suggests that, at least for these benchmark points, the bag model provides a reasonable approximation to the fluid profiles obtained from the full equation of state.
However, if the model under consideration contains particles with masses close to the characteristic temperature of the phase transition and large relativistic degrees of freedom, one still can expect significant deviations in the speed of sound, as well as noticeable deviations in the gravitational wave spectrum.

Next, we vary the method of computing the kinetic energy fraction $K$.
In the literature, it is common practice to use fitting formulas Eq.(\ref{eq: fitting deflagration}) - Eq.(\ref{eq: fitting detonation}) to compute the kinetic energy fraction $K$, which allows for a quick estimation of the resulting gravitational wave spectrum --- an approach that, in effect, maps the model under consideration onto the bag model.
We compare the fitted $K$ with that obtained through fluid profile integration, and the results as shown in Fig.\ref{fig:bag_K_Omega} of the Appendix.
As can be seen, the fitting formula for the deflagration mode tends to overestimate the value of $K$, leading to an overestimation of the peak amplitude of the gravitational wave spectrum. However, for the detonation mode, the fitting formula is consistent with the direct calculations well. Across the entire scanned parameter space, the average deviation in the peak value of gravitational wave spectrum is found to be approximately 14.3\% (deflagration) and 2.9\% (detonation), respectively.

Then, we vary the bubble wall velocity while keeping all other inputs — the EOS and the method for computing $K$ — unchanged. For the choice of EOS, the full EOS is adopted in this work, as comparisons with the bag model have been thoroughly explored in previous studies. The two scenarios being compared are:
\begin{itemize}
    \item a fixed wall velocity of $v_w = 0.3$ for deflagration and $v_w = 0.9$ for detonation;
    \item a velocity obtained from the LET approach.
\end{itemize}

\begin{figure}
    \centering
    \includegraphics[width=\linewidth]{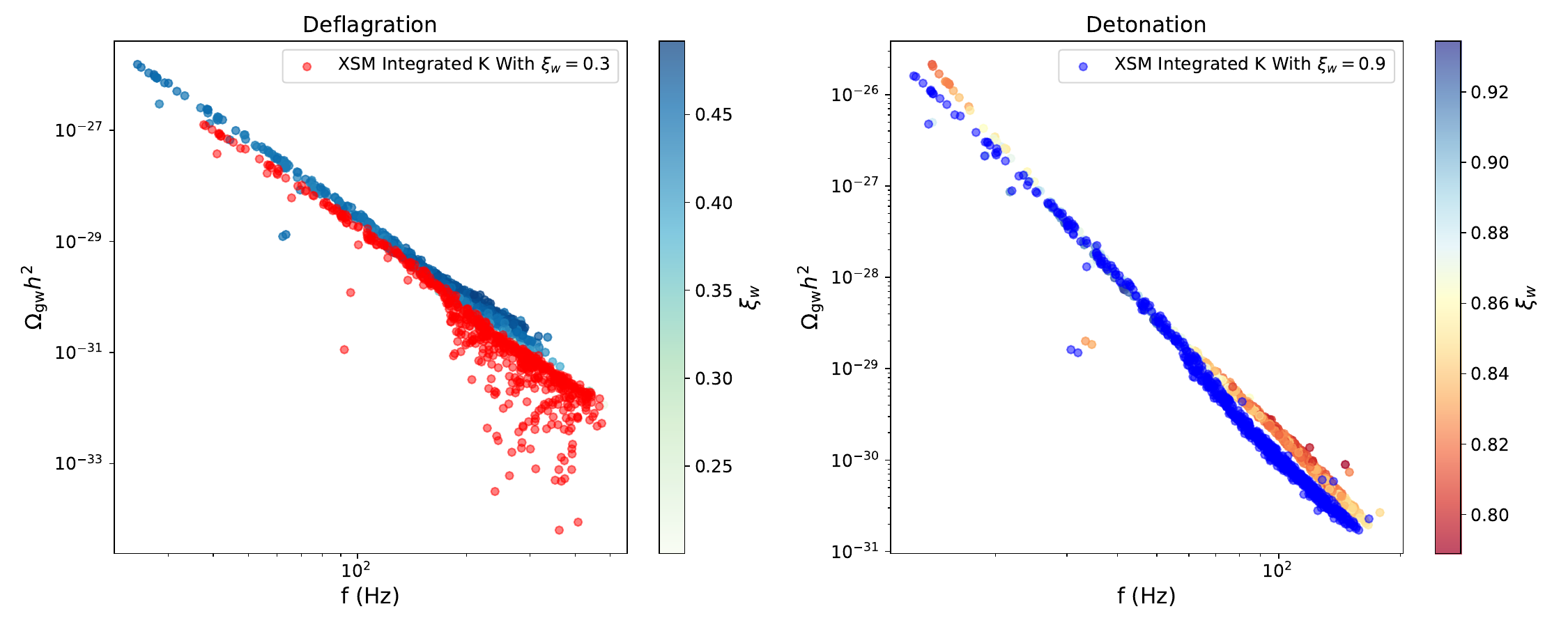}
    \includegraphics[width=\linewidth]{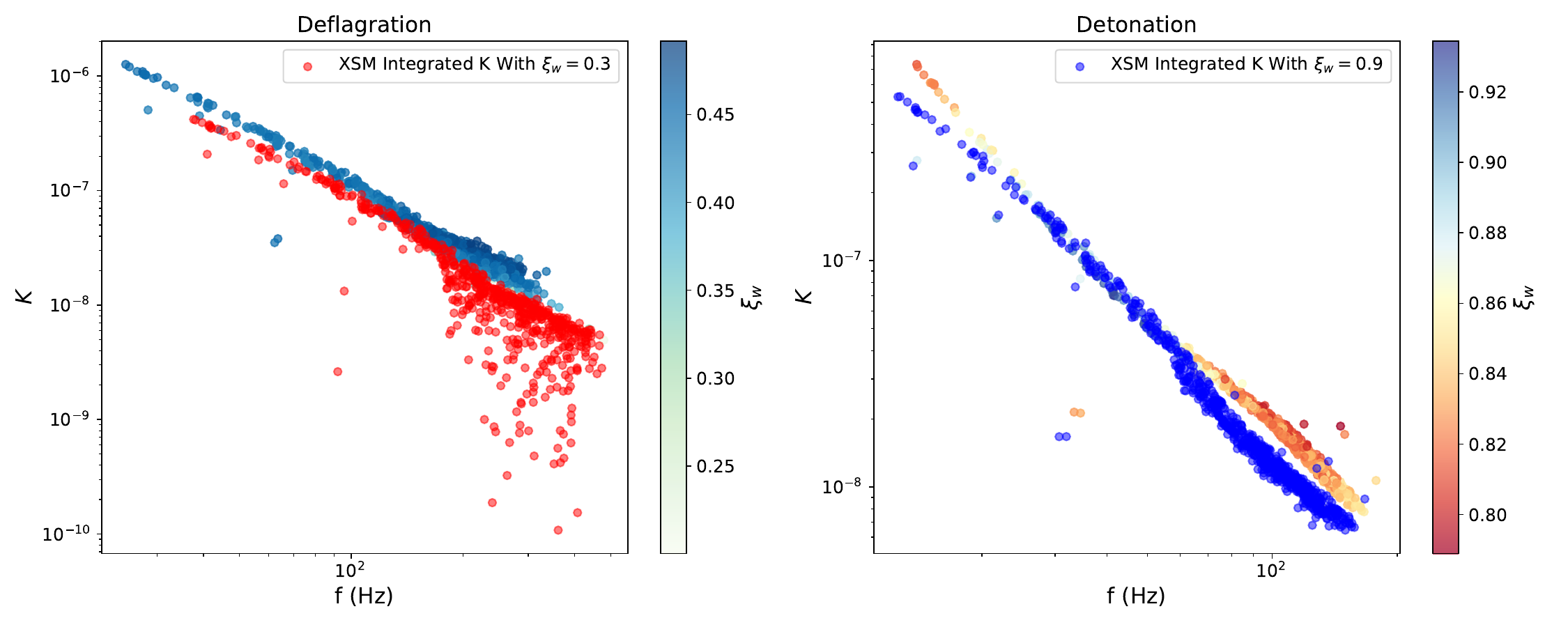}
    \caption{Comparison of gravitational wave peak spectra and energy budgets obtained by fixing the EOS and the method for calculating the kinetic energy fraction $K$, while varying the bubble wall velocity $\xi_w$. The left panel corresponds to deflagration modes, and the right panel to detonation modes. Red points represent results obtained with fixed wall velocities, whereas points colored by the color bar correspond to calculations using wall velocities derived from the LTE approximation.
}
    \label{fig:different_vw}
\end{figure}

The resulting spectra are shown in Fig.\ref{fig:different_vw}. In the case of detonation, except for a small set of parameter points whose peak frequencies lie between 20 Hz and 60 Hz, the fixed-velocity scenario yields a spectrum amplitude lower than that obtained from the LET result, regardless of whether the fixed $\xi_w$ is larger or smaller than the LET-derived value. 
However, in the deflagration case, the fixed-velocity results in the lower-right region fluctuate strongly and differ markedly from those based on the LET approach.
To further investigate the source of the discrepancy, we extract the relevant parameter points and first rule out the possibility that caused by deviations in the speed of sound. 
We suspect that the fluid profile is highly sensitive to the choice of wall velocity and impacts the kinetic energy fraction $K$, resulting in considerable deviations in the gravitational wave peak amplitude. 
We plot $K$ and indeed observe features similar to those seen in the gravitational wave peak amplitude. In the region of significant deviation, the fluid profiles corresponding to the two velocity choices are qualitatively similar in shape, but their peak values differ by orders of magnitude.

\begin{figure}[tb]
	\centering
        \includegraphics[width=1\textwidth]{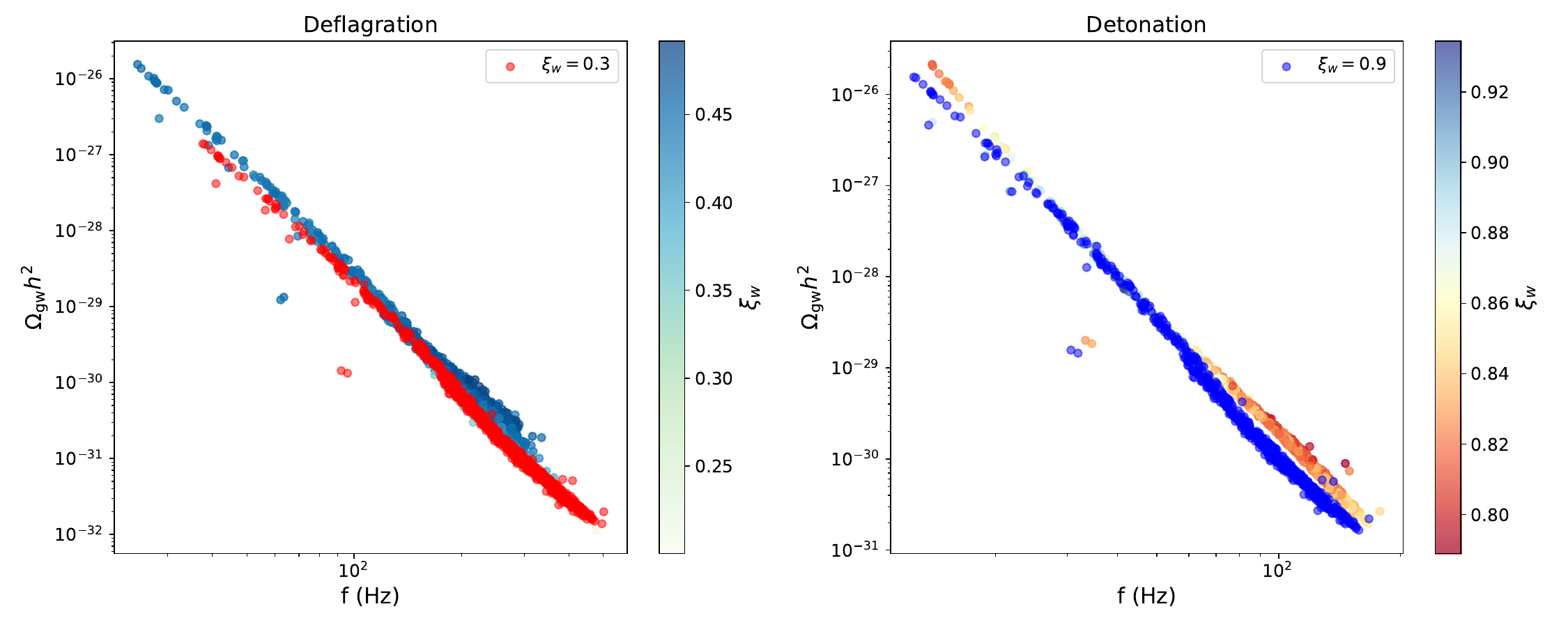}
	  \caption{Gravitational wave predictions comparing different definitions of the kinetic energy fraction $K$ and equation of state models. (Left): GW peak amplitudes vs. peak frequencies for deflagration solutions.  (Right): Same for detonation. The color bar represents the results obtained using the most precise computational method. Pure red (blue) points represent predictions using the bag model with fitted $K$ and fixed wall velocities ($\xi_w = 0.3$ for deflagration, $\xi_w = 0.9$ for detonation).}
	\label{GW Scan results for the singlet-extension of SM}
\end{figure}

Finally, we systematically compare the gravitational wave spectra obtained from the most refined calculation with those from the most simplified setup. The refined calculation uses the full EOS, computes the efficiency factor $K$ via hydrodynamic integration, and determines the bubble wall velocity $\xi_w$ using the LTE approximation for both deflagration and detonation modes. In contrast, the simplified setup adopts the bag model EOS, calculates $K$ using a fitting formula, and fixes the wall velocity to $\xi_w = 0.3$ and $\xi_w = 0.9$ for deflagration and detonation, respectively. The resulting spectra are shown in Fig.\ref{GW Scan results for the singlet-extension of SM}, illustrating the discrepancy caused by varying assumptions about $\xi_w$ and $K$. The color bar indicates the wall velocity derived from hydrodynamic calculations, while red (blue) points represent results using the fitted $K$ with fixed $\xi_w=0.3$ ($\xi_w=0.9$).
Within the scanned parameter space, we find that when the precisely calculated bubble wall velocity closely matches the fixed value used in the fitting formulas, the predicted gravitational wave peak amplitudes under both fluid motion regimes remain largely consistent, indicating a degree of universality in the expression for $K$.
However, significant deviations appear when the two velocities differ.
In the deflagration regime, the gravitational wave peak amplitude increases with the wall velocity. In contrast, the detonation regime behaves differently: the peak is maximized around $\xi_w \approx 0.8$, suggesting that the gravitational wave amplitude is not simply proportional to the wall velocity.
This is because, while the gravitational wave signal scales positively with $\xi_w$, it inversely depends on $\beta/H$. In regions where $\xi_w\approx0.8$, larger gravitational wave amplitudes correspond to smaller $\beta/H$ values, enhancing the signal despite similar wall velocities.
Finally, we calculated the relative errors of both the peak frequency and the peak amplitude. In the deflagration regime, the average relative error of the peak frequency $f_{\rm{deflagration}}$ reaches approximately 48\%, while the error in the peak amplitude $\Omega_{\rm{sw}} h^2_{\rm{deflagration}}$ is around 90\%. The detonation regime shows much smaller discrepancies, with average relative errors of about 6\% for $f_{\rm{detonation}}$ and 18\% for $\Omega_{\rm{sw}} h^2_{\rm{detonation}}$. 


Comparing the results of Fig.\ref{fig:different_vw} and Fig.\ref{GW Scan results for the singlet-extension of SM}, we find that the variation in the final gravitational wave spectrum is almost entirely determined by the bubble wall velocity. In the case of deflagration, if the bubble wall velocity is treated as an input parameter, mapping the model to the bag model is a more appropriate approach. In particular, the gravitational wave results based on the fitted $K$ align more closely with the most refined results.

\section{Conclusion}\label{sec5}
Gravitational waves from first-order phase transitions offer a promising probe of new physics, but their predictions are highly sensitive to the bubble wall velocity. 
To reduce such uncertainties, we propose a computational framework based on the LTE assumption to determine the stable bubble wall velocity and the corresponding energy budget, using the full one-loop finite-temperature effective potential. The reliability of this approach is validated through comparison with results obtained from bag model. We also use \texttt{WallGo} to crosscheck our results. Taking the xSM as an example, we first investigated the distribution of the different fluid motion patterns in the space of our chosen parameters, and then estimate the uncertainty of the gravitational wave spectra obtained via different ways. Our results show that:
\begin{itemize}
    \item Deflagration is the most likely fluid motion mode to occur. 
    \item Within the scanned parameter space, the bag model serves as a reliable approximation to the full EOS.
    \item When the precisely calculated wall velocity is close to the fixed value, the fitted formulas for efficiency factor yield consistent results. However, when there is a significant mismatch between them, deviations arise—especially in the deflagration regime. In the detonation regime, the relative errors in the predicted peak frequency and gravitational wave amplitude can reach up to 6\% and 18\%, respectively. The deflagration regime shows much larger discrepancies, with relative errors as high as 48\% for the peak frequency and 90\% for the gravitational wave amplitude. 
    \item In the deflagration scenario, when the bubble-wall velocity is taken as an external input, the gravitational wave spectrum computed using the bag model more closely aligns with the LET-based result than the spectrum obtained from the full EOS.
\end{itemize}
We hope that our work can contribute to the precise detection of gravitational wave spectra in the future. Furthermore, more accurate comparisons require solving the Boltzmann equations to obtain a more precise determination of the bubble wall velocity. We will systematically quantify the above uncertainties with greater precision in our future studies.

\section{Appendix}
\subsection{Code}
In the computational framework we have developed, users can compute the bubble wall velocity for a given BSM model by importing the effective potential they have implemented in the following way:
\begin{lstlisting}
from BSMmodel import realmodel
\end{lstlisting}
Here, \texttt{BSMmodel} refers to the Python file that implements the effective potential using the \texttt{CosmoTransitions} package, and \texttt{realmodel} is the user-defined model class within our framework. During class instantiation, the following parameters need to be provided by the user and passed in to initialize member variables and may be used in subsequent function calls to configure the model further:
\begin{lstlisting}
class Hydrodynamics_profile:
    def __init__(self, T_ref, T_C, model_parameters, low_phase_key, high_phase_key, gs_phase_key):
         self.model = realmodel(model_parameters[0], model_parameters[1], model_parameters[2])
         self.phase_information = self.model.getPhases()
\end{lstlisting}
\begin{itemize}[label=\textbullet, left=0pt] 
\item \texttt{T\_ref} is a reference temperature, which can be chosen by the user depending on the context, such as the nucleation temperature or the percolation temperature.
\item \texttt{T\_C} represents the critical temperature and is used in the deflagration mode to provide a reasonable initial guess for $T_-$ during the solution of initial conditions.
\item To support multi-step phase transitions, \texttt{low\_phase\_key} and \texttt{high\_phase\_key} indicate the phase indices corresponding to the true vacuum and false vacuum in each transition step, respectively.
\item \texttt{gs\_phase\_key} represents the index of the phase with the lowest energy density at zero temperature (i.e., the ground state).
\item The \texttt{mode\_parameters} argument contains the input parameters required to configure the effective potential of the underlying BSM model, such as masses or vacuum expectation values.
\end{itemize}
Within this framework, several parameters are defined in the computational framework to specify the number of sampling points used for calculating bubble wall velocities under different hydrodynamic modes, (e.g., \texttt{self.hybrid\_cal\_T\_plus\_num=10000} etc.). Increasing the number of points generally improves accuracy by better resolving the numerical solution. In this work, the sampling densities are set sufficiently high to ensure numerical convergence, making the results reliable within the required precision. Users may adjust these settings as needed for their desired accuracy.
Once initialized, the phase structure is automatically extracted via \texttt{self.model.getPhases()} for use in the wall velocity computation.
\begin{lstlisting}
    def P_density(self, X, T):
        return -self.model.Vtot(X, T)

    def E_density(self, X, T):
        dVdT = (self.model.Vtot(X, T + self.delta_T)
                - self.model.Vtot(X, T - self.delta_T)) / (2 * self.delta_T)
        return self.model.Vtot(X, T) - T * dVdT

    def W_density(self, X, T):
        return self.P_density(X, T) + self.E_density(X, T)
\end{lstlisting} 
The functions \texttt{P\_density(self, X, T)}, \texttt{E\_density(self, X, T)}, and \texttt{W\_density(self, X, T)} correspond to the thermodynamic quantities defined in Eqs.~\eqref{eq: state parameter relation p}--\eqref{eq: state parameter relation w}, namely the pressure, energy density, and enthalpy density, respectively.
\begin{lstlisting}
    def mu(self, xi_w, v): 
        v = (xi_w - v) / (1 - xi_w * v)
        if (v > 0) and (v < 1):
            return v
        raise ValueError('mu is wrong!')
        
    def cs_squ(self, phase_key, T):
        X = self.phase_information[phase_key].valAt(T)
        cs_seq=(self.P_density(X, T + self.delta_T) - self.P_density(X, T - self.delta_T)) / (
                self.E_density(X, T + self.delta_T) - self.E_density(X, T - self.delta_T)+1e-5)
        return cs_seq
\end{lstlisting} 
The method \texttt{def mu(self, xi\_w, v)}, as defined in Eq.~\eqref{Lorentz velocity transformation}, represents the Lorentz transformation between the bubble-wall frame and the bubble-center frame. The function \texttt{def cs\_squ(self, phase\_key, T)} returns the square of the speed of sound, as defined in Eq.~\eqref{eq: speed of sound}. 
 \begin{lstlisting}
    def find_profiles(self):
\end{lstlisting} 
The method \texttt{find\_profiles(self)} is responsible for computing the Jouguet velocity as well as the three types of fluid configurations. By default, fluid profile data is saved in \texttt{.npy} format, with the option for users to disable saving to reduce storage usage. It is also possible to compute only one or a subset of the available fluid configurations, depending on the user's needs.
\begin{lstlisting}
if __name__ == '__main__':
    model = Hydrodynamics_profile(T_ref, T_C, model_parameters, low_phase_key, high_phase_key, gs_phase_key)
    model.find_profiles()
\end{lstlisting} 
In the main program, an instance of the \texttt{Hydrodynamics\_profile} class is created using the previously defined parameters. The \texttt{model.find\_profiles()} method is then called to compute the required results.

\subsection{The results for bag model}
\begin{figure}
    \centering
    \includegraphics[width=\linewidth]{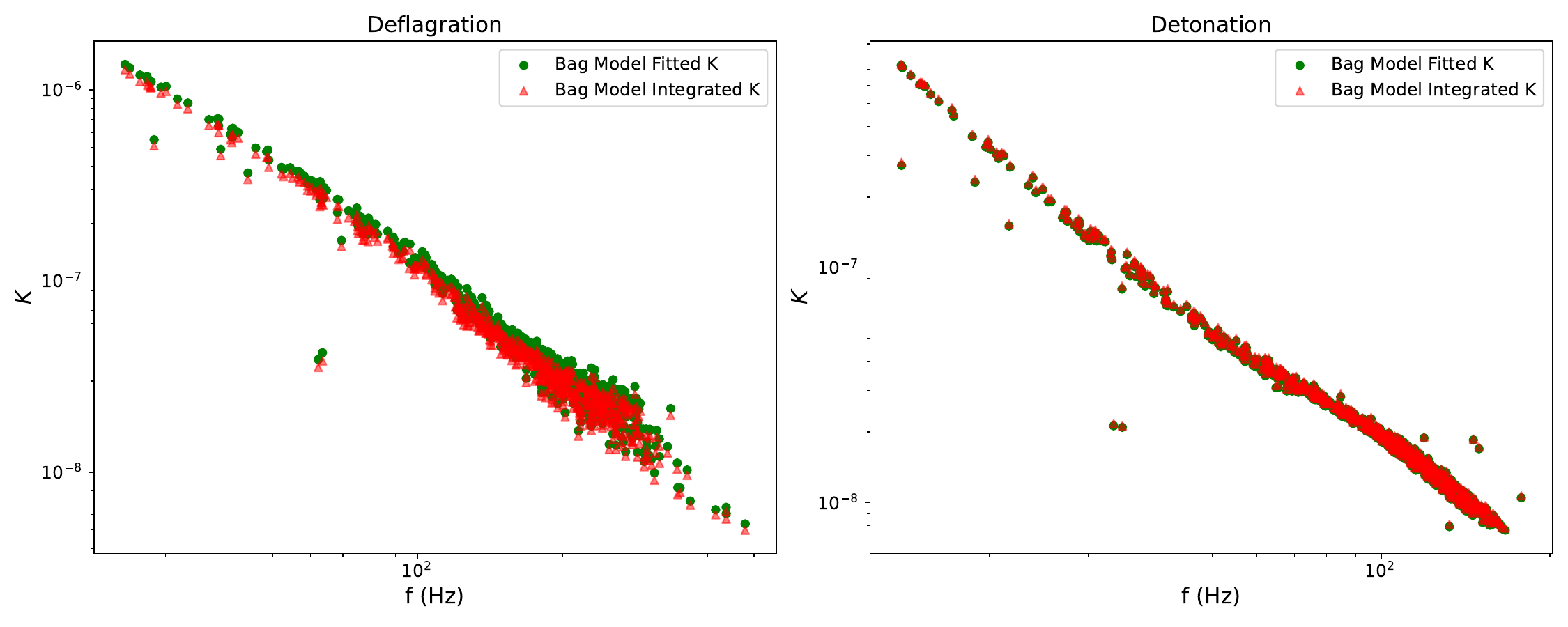}
     \includegraphics[width=\linewidth]{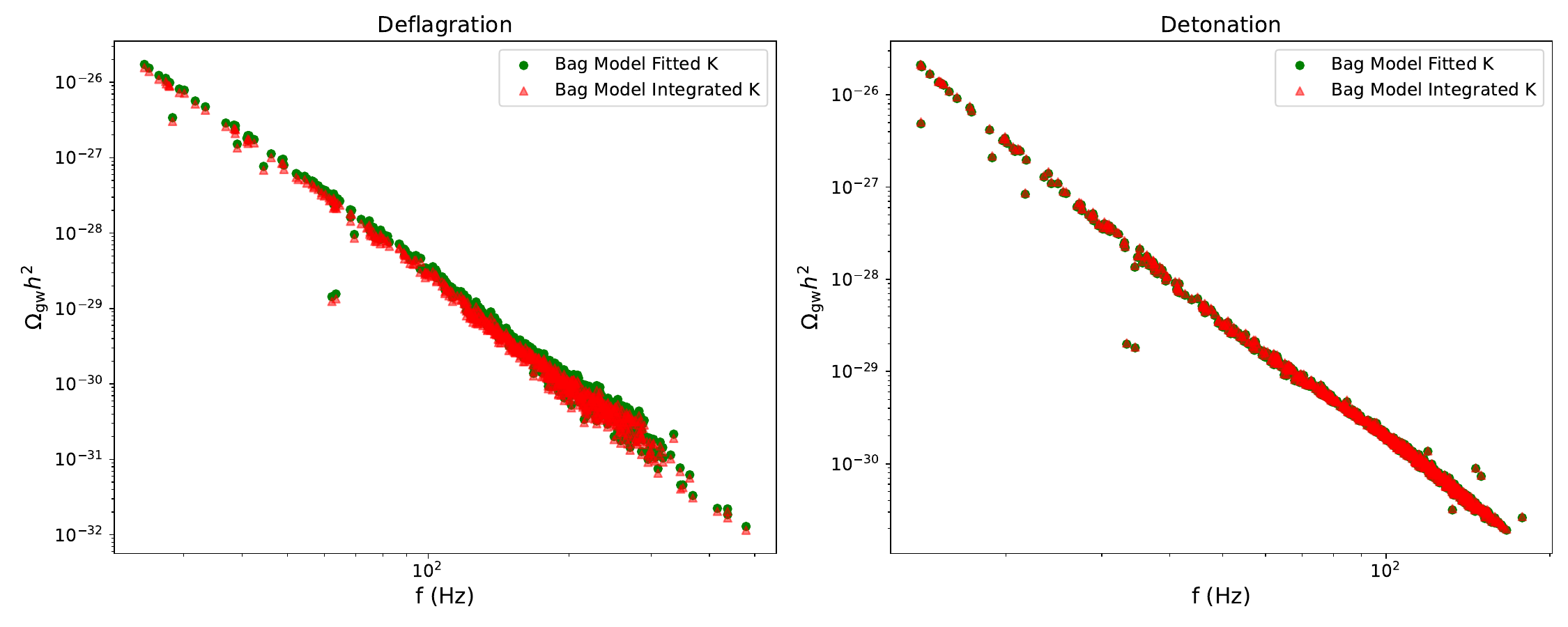}
    \caption{Comparison of gravitational wave peak spectra and energy budgets resulting from different calculations of 
    $K$ within the bag model framework. The left panel corresponds to deflagration modes, and the right panel to detonation modes. Red triangles data points represent results obtained using the integral expression Eq.(\ref{eq:the kinetic energy fraction}), while green circular data points correspond to those calculated using the fitting formula Eq.(\ref{eq:the kinetic energy fraction fit}).}
    \label{fig:bag_K_Omega}
\end{figure}

Here, we present the relevant results obtained using the bag model in Fig.\ref{fig:bag_K_Omega}.

\newpage
\section*{Acknowledgments}
We would like to thank Peter Athron for helpful discussions. This work is supported by the National Natural Science Foundation of China (No. 11975013, 12335005 and 12105248) and by the Projects No. ZR2024MA001 supported by Shandong Provincial Natural Science Foundation.

\section*{Note added:}
Towards the completion of our work, we became aware of a related study by Marcela Carena et al.~\cite{Carena:2025flp}. In that work, the authors employed the LTE approximation to examine how the bubble wall velocity depends on the parameters $\alpha$ and $v/T$ in two different models. They also investigated the corresponding gravitational wave spectra and the implications for baryogenesis within the same frameworks. While our methodology is broadly consistent with theirs, our study is primarily focused on the quantitative uncertainties in gravitational wave predictions, particularly those arising from the choice of the effective potential and the specification of the bubble wall velocity. A more refined analysis would require a full three-dimensional effective field theory treatment, along with solving the Boltzmann equations to determine the bubble wall velocity more accurately. We plan to address these aspects in future work.

\bibliographystyle{JHEP}
\bibliography{ref}
\end{document}